\documentclass[a4paper,fleqn,usenatbib]{mnras}
\usepackage{hyperref}
\usepackage{subfig}
\usepackage{subfloat}
\usepackage{bm}
\usepackage{graphicx}
\usepackage[T1]{fontenc}
\usepackage{ae,aecompl}

\title[Numerical predictions for planets in the debris discs of HD~202628 and HD~207129]{Numerical predictions for planets in the debris discs of HD~202628 and HD~207129}
\author[E. Thilliez et al.]{E. Thilliez$^{1}$\thanks{e-mail: ethilliez@astro.swin.edu.au}, S. T. Maddison$^1$\\
$^1$Centre for Astrophysics and Supercomputing, Swinburne University of Technology, Hawthorn, VIC 3122, Australia}

\date{Released 2015 Xxxxx XX}

\pubyear{2015}

\begin{document}
\label{firstpage}
\pagerange{\pageref{firstpage}--\pageref{lastpage}}
\maketitle

\begin{abstract}
Resolved debris disc images can exhibit a range of radial and azimuthal structures, including gaps and rings, which can result from planetary companions shaping the disc by their gravitational influence. Currently there are no tools available to determine the architecture of potential companions from disc observations. Recent work by ~\cite{2014ApJ...780...65R} presents how one can estimate the maximum mass and minimum semi major axis of a hidden planet empirically from the width of the disc in scattered light. In this work, we use the predictions of Rodigas et al. applied to two debris discs HD 202628 and HD 207129. We aim to test if the predicted orbits of the planets can explain the  features of their debris disc, such as eccentricity and sharp inner edge. We first run dynamical simulations using the predicted planetary parameters of Rodigas et al., and then numerically search for better parameters. Using a modified N-body code including radiation forces, we perform simulations over a broad range of planet parameters and compare synthetics images from our simulations to the observations. We find that the observational features of HD~202628 can be reproduced with a planet five times smaller than expected, located 30~AU beyond the predicted value, while the best match for HD~207129 is for a planet located 5-10 AU beyond the predicted location with a smaller eccentricity. We conclude that the predictions of Rodigas et al. provide a good starting point but should be complemented by numerical simulations.
\end{abstract}

\section{Introduction}
Many main sequence stars are known to host dusty debris discs, which can be accompanied by planetary companions. The small dust grains, traced by scattered light images, experience radiation forces from the host star, such as radiation pressure and Poynting-Robertson drag ~\citep{1979Icar...40....1B} which considerably shorten their survival time. However debris disks with ages $\ge$ 10 Gyr are observed ~\citep{2012A&A...548A..86L}, implying that an unseen population of planetesimals must be evolving through collisions and fragmentation, continually replenishing the disk with new dust. Planets can potentially stir the parent body belt to replenish the discs with new dust, and are therefore a key element for debris disc survival ~\citep{2009MNRAS.399.1403M}. 

Amongst the 40 currently known resolved debris discs, recent images show both radial and azimuthal structures, such as gaps, eccentric rings and warps. While local pressure bumps -- created, for example, by the photoelectric instability effect \citep{2013Natur.499..184L} -- can induce similar asymmetries, planetary companions can produce all these features through their gravitational influence  ~\citep{2012A&A...547A..92T,2015arXiv150607187N}. \cite{1999ApJ...527..918W} developed the secular perturbation theory for debris discs and suggested that an eccentric planet could cause the double-ringed disc of HD~107146 \citep{2015MNRAS.453.3329P}.  ~\cite{2012A&A...542A..18L} suggested that the warp of the $\beta$ Pictoris disc could result from the disc interacting with the planet $\beta$ Pic b, which is supported by the numerical model developed by \cite{2015arXiv150607187N}. 

While the impact of the presence of one planet on the structure of a debris disc has been well studied and is now understood, it remains difficult to constrain the planet's orbital parameters from debris discs observations. This issue was recently addressed for the first time by ~\cite{2014ApJ...780...65R}, who derived an empirical formula linking the width of a debris disc in scattered light and an estimate of planetary parameters.

 ~\cite{2014ApJ...780...65R} applied their empirical formula to provide a rough estimation of the potential location and mass of a potential planetary companion for HD~202628 and HD~207129. Both are G type stars hosting discs with scattered light features suspected of being shaped by an unseen planet, such as an eccentric ring for HD~202628 \citep{2012AJ....144...45K} and a very sharp inner edge for HD~207129 \citep{2010AJ....140.1051K}.

We aim here to test if the predicted orbits of the planets can indeed explain the peculiar features of their debris disc by running dynamical simulations of these two systems. In order to model a debris disc interacting with a planet, we have developed a modified N-body code that takes into account radiation forces acting on the small grains in debris discs that are traced by scattered light. Using the planet parameters estimated from ~\cite{2014ApJ...780...65R} and observed disc parameters from ~\cite{2010AJ....140.1051K,2012AJ....144...45K} as initial conditions, we first dynamically model the systems and compare our results with observations using the radiative transfer code \textit{MCFOST} to create synthetic observations from our simulation results. We then explore a broader range of initial planet parameters to better fit to the observation as required.

 The paper is organized as follows: first, we present the characteristics of HD~202628 and HD~207129 in Section 2 before expanding on the predictions of ~\cite{2014ApJ...780...65R} for their potential planetary companions in Section 3. Section 4 describes the numerical method we use, and finally we report our findings in Section 5 and discuss the results in the Section 6.

\section{HD~202628 and HD~207129}
HD~202628 is a G2 type star at a distance of 24.4~pc ~\citep{2010ApJ...710L..26K}, with an age of $\sim 2.3$~Gyr. Its debris disc was first imaged in scattered light by ~\cite{2012AJ....144...45K} using the \textit{Hubble Space Telescope} (HST). It showed a belt slightly narrower than our Kuiper Belt with an observed width to mean radius ratio $\Delta r$/$r_{0}$ $\sim$ 0.4 (cf 0.6 for the Kuiper Belt),  where the width, $\Delta r$, is defined as the full width half maximum (FWHM) of the surface brightness profile and the mean radius, $r_{0}$, is the peak location in the profile. Among its features, this inclined debris disc ($i=61^{\circ}$) exhibits a large offset of about 28 AU (deprojected) from the star and a very sharp inner edge with a radial brightness profile in $r^{\alpha}$ with $\alpha=12$ as measured by ~\cite{2012AJ....144...45K}. In the projected image, the inner edge is fit by an ellipse of eccentricity 0.18 and semi-major axis 158 AU, and the disc is roughly located between 150 and 220 AU. The authors suggested that a planet of 10 M$_{\rm J}$ at 120 AU on an eccentric orbit could be responsible for the offset and eccentric disc, while recent infrared images taken with \textit{Herschel/PACS} suggests a planet at 100 AU ~\citep{2013AAS...22114414S}. By applying their derivation of the gap law (linking the width of the gap to the mass of the planet opening the gap) and assuming a 15 M$_{\rm J}$ planet, \cite{2015ApJ...798...83N} estimate the potential planet to lie between $86< a_{p} <158$~AU, while \cite{2014MNRAS.443.2541P}, based on their derivation of the Hill radius, suggest $80<a_{p}<130$~AU. The main parameters of HD~202628 are summarized in Table ~\ref{Table1}. \\

HD~207129 is a G0 type star at a distance of 16~pc ~\citep{1997A&A...323L..49P} with an estimated age of about 1 Gyr, whose  debris disc was first observed by ~\cite{1999A&A...350..875J} with the \textit{Infrared Space Observatory}, and more recently by ~\cite{2010AJ....140.1051K} in scattered light and near-infrared with \textit{HST/ACS} and \textit{Spitzer/MIPS}. Other observations over a broad range of wavelengths from 9 to 870 $\mu$m are available -- and summarized in \citep{2011A&A...529A.117M} --  as well as resolved \textit{Herschel/PACS} data from ~\cite{2012A&A...537A.110L}. ~\cite{2010AJ....140.1051K} fit the scattered light observations and found an inclined disc ($i=60^{\circ}$) with width 30 AU located at 163 AU, making HD~207129 one of the narrowest ring detected with $\Delta r$/$r_{0}$ $\sim$ 0.2. The spectral energy distribution (SED) of this well-observed debris disk exhibits steepy rising excess emission beyond 30 $\mu$m and faint emission interior to 148 AU on the scattered light image, which is interpreted as a very sharp inner disk, a common signature of the presence of a planet. ~\cite{2010AJ....140.1051K} noticed a bright point source near the south edge of the coronographic image (see their Figure 1) but were unable to determine its comotion with the system. Such an object is estimated to have a mass of $\sim$ 20 M$_{\rm J}$ ~\citep{2010AJ....140.1051K}, and we estimate its distance from the star to be about $7^{''}$, i.e. 110 AU (projected). The main parameters of HD~207129 are summarized in Table~\ref{Table1}.
\begin{table}
\renewcommand{\arraystretch}{1.0}
\caption{Properties for HD~202628 from Krist et al. (2012) and for HD~207129 from Krist et al. (2010).}
\label{Table1}
\centering
\begin{tabular}{ccc}
\hline
Properties &  HD~202628 & HD~207129 \\
\hline
& Stellar properties & \\
\hline
Spectral type & G2 V & G0 V \\
Age & 2.3 $\pm$ 1 Gyr & $\sim$1~Gyr\\
Luminosity, $L_{\ast}$ & $\sim$ 1~$L_{\odot}$ & 1.2~$L_{\odot}$\\
Mass, $M_{\ast}$ &  $\sim$ $1~M_{\odot}$ & $\sim$ $1~M_{\odot}$  \\
Distance, $d$ & 24.4~pc$^{b}$ & 16.0~pc$^{c}$\\
\hline
& Disc properties & \\
\hline
Width, $\Delta r^{a}$ &  150--220~AU & 148--178~AU\\
Mean radius, $r_{0}$ & $\sim$ 182 AU & 163 AU \\
Disc width ratio, $\Delta r$/$r_{0}$ & 0.4 & 0.2 \\
Eccentricity, $e$ & 0.18 $\pm$ 0.02 & <0.08$^{d}$\\
Deproj. disc offset, $\delta$ & 28.4 AU & $<13.5$~AU$^{e}$\\
Inner edge power law fit, $r^{\alpha}$ &   $r^{12}$    &   unknown  \\
Line-of-sight inclination, $i$ & 61$^{\circ}$ $^{a}$ &  60$^{\circ} \pm$3 \\
Position Angle, $PA$ & 134$^{\circ}$ $^{a}$ & 127$^{\circ}$ $\pm$3 \\
\hline
\end{tabular}
\\
\small{$^{a}$ Best fits  ~\cite{2014ApJ...780...65R}, $^{b}$ \cite{2010ApJ...710L..26K},$^{c}$ \cite{1997A&A...323L..49P},\\
 $^{d}$  ~\cite{2014ApJ...780...65R}, $^{e}$ estimated from the 7.3 AU projected offset of ~\cite{2014ApJ...780...65R}.}
\end{table}

\section{Empirical predictions from Rodigas et al. (2014)}
While theoretical models such as the secular perturbation theory of \cite{1999ApJ...527..918W} and numerical simulations ~\citep{2002ApJ...578L.149Q,2005ApJ...625..398D,2010ApJ...717.1123M} can determine the impact of a massive planet on the structure of a debris disc, it is currently difficult to extrapolate and constrain the orbit of a potential planetary companion from debris disc observations. This issue was recently tackled by ~\cite{2014ApJ...780...65R} in the case of a single interior planet interacting with a debris disc, potentially causing the disc to be offset with a sharp inner edge. By performing 160 unique N-body simulations covering a broad range of planetary masses, initial disc and planet eccentricities and radiation force intensities, the authors found a correlation between the final width of the debris disc in scattered light, $\Delta r$, and the mass of the planet, $m_{p}$. Fitting this relation, the authors provide an empirical formula where observers can input the FWHM of the deprojected surface brightness profile from any scattered light imaged debris disc to obtain the maximum mass of the planet shaping the disk (see their Equation 5). \\

By studying the migration of a planet interacting with a massive planetesimal belt located exterior to the planet, ~\cite{2014ApJ...780...65R} also derived an expression linking the final location of the planet, the observed position of the inner edge of the disc and the mass of the planet. With this relation, an observer can obtain a minimal semi-major axis for the shepherding planet (see their Equation 2).

The authors  present several applications of these empirical equations for well-observed systems in scattered light, including HD~202628 and HD~207129. Firstly, using the scattered light surface brightness profile of HD~202628 from \cite{2012AJ....144...45K}, they derived a disc width ratio $\Delta r/r_{0} \sim$ 0.4 for the disc, suggesting a perturbing planet of mass $\sim$ 15~M$_{\rm J}$ located at a semi-major axis $a_{p} > 71$~AU. For HD~207129, they used the \textit{HST} image of ~\cite{2010AJ....140.1051K} and obtained a disc width ratio of 0.18, suggesting a perturbing planet of mass $\sim$ 4~M$_{\rm J}$ beyond 92~AU. The predicted planet masses and orbits from ~\cite{2014ApJ...780...65R} for these two systems are summarized in Table~\ref{Table2}. We will use these predicted values as starting values for our simulations.
\begin{table}
\renewcommand{\arraystretch}{1.0}
\caption{Predicted masses and orbits for potential planets in the debris disc of HD~202628 and HD~207129 from Rodigas et al. (2014).}
\label{Table2}
\centering
\begin{tabular}{cccc}
\hline
Star & $m_{p}$ (M$_{\rm J}$) & $a_{p}$ (AU) & $e_{p}$ \\
\hline
HD~202628 & 15.4$\pm$ 5.5  & $>$71  & 0.18\\
HD~207129 &  4.2$\pm$ 2.3 & $>$92  & $<$0.08\\
\hline
\end{tabular}
\end{table}

\section{Method}
In this section we describe the modified N-body integrator used to model HD~202628 and HD~207129, the simulation suite then we run, and the radiative transfer code used to compare with \textit{HST} scattered light images.
\subsection{The integrator}
Our modified N-body integrator uses the regular mixed variable sympletic (RMVS) integrator in the software package \textit{SWIFT}, developed by ~\cite{1994Icar..108...18L}. For grains in the disc with sizes, $s$, between $10^{-6}$-$10^{-2}$ m, forces other than gravity can have a significant impact, specifically radiation pressure from the central star. A common way to express the radiation pressure is to equate it to a fraction, $\beta$, of the gravitational force between the central star and the grain, where $\beta$ decreases linearly with grain size. Thus, if we use common units and assuming the grains to be a blackbody, $\beta$ can be simply expressed by:
\begin{equation}
\beta=0.577 \frac{L_{\ast}}{\rm{L}_{\odot}} \left(\frac{\rho}{\rm{g/cm}^{3}}\right)^{-1} \left(\frac{s}{\mu\rm{m}}\right)^{-1} \left(\frac{M_{*}}{\rm{M}_{\odot}}\right)^{-1},
\label{eq:1}
\end{equation}
where $L_{\ast}$ is the luminosity of the star, $\rho$ is the density of the grains and $M_{\ast}$ the stellar mass. Another force which can impact on the particle's orbit is the stellar wind. This results from the impact of stellar protons on the grains, which leads to mass loss. As sputtered molecules from the grain carrying momentum away, the orbit of the grains decreases toward the star. ~\cite{1979Icar...40....1B} derived the ratio, $sw$, between the stellar wind and radiation pressure and found it independent of the grain size for $s$ $> 1~\mu$m. In this study, we will use the value of $sw$ equal to 0.05\footnote{The most common value used in the literature is $sw$=0.35, which represents a magnetite grain ($\rho$ $\sim$ 5~g/cm$^{3}$) experiencing a solar type wind. However since ~\cite{2010AJ....140.1051K} assumed a density of 2.5~g/cm$^{3}$ for their silicate grains, we chose the stellar wind coefficient for obsidian grains which have a density $\sim$ 2.5~g/cm$^{3}$.}, which corresponds to a solar type wind for an obsidian grain ~\citep{1982A&A...107...97M}, appropriate for both HD~202628 and HD~207129 which are G type stars. 

The total acceleration on a grain due to radiation forces, truncated at the second order, is given by:
\begin{equation}
\frac{d^{2} \bm{r}}{dt^{2}}=F_{\rm grav}\left(\beta \bm{r} -\frac{\beta(1+sw)}{c}(v_{r}\bm{r}+\bm{v})\right),
\label{eq:2}
\end{equation}
where \textbf{\textit{r}}, \textbf{\textit{v}} are the position and velocity vectors, $\beta$ is the ratio $F_{\rm rad}/F_{\rm grav}$, and $sw$ is the ratio of solar wind drag to radiation pressure. Since there is no general solution for the motion of a particle under the influence of radiation and gravitation forces in a system comprising a star and a planet, we followed the approach of \cite{2005ApJ...625..398D} and tested our code by comparing simple two- and three-body simulations with analytic solutions derived by ~\cite{2002AJ....124.2305M} and ~\cite{1997Icar..128..354L}. 
\subsection{Simulations}
To numerically test the predictions of ~\cite{2014ApJ...780...65R}, we first run simulations of the HD~202628 and HD~207129 systems using the initial conditions for the planet provided by the predicted values from Rodigas et al. (see Table~\ref{Table2}) and initial conditions for the debris discs from the observations of ~\cite{2010AJ....140.1051K,2012AJ....144...45K} (see Table~\ref{Table1}). We then create a synthetic scattered light image at 1.03 $\mu$m of the disc to compare with the \textit{HST} observations of ~\cite{2010AJ....140.1051K,2012AJ....144...45K}. If our results do not well match the observed features of the debris discs, we run additional simulations exploring a range of planetary parameters to determine a better fit to the observational data.

To model the debris discs, we include the central star, a massive planet and a swarm of massless test particles that represent dust grains released by a parent body belt of planetesimals.
To model the debris disc, we would ideally like to use a large number of parent bodies, however because test particles can be removed from the system by two separate physical processes, Poynting-Robertson (PR) drag and radiation pressure, we instead model the continuous release of test particles resulting from parent body collisions using the following stacking method. During the simulations, after a period of $t=t_{init}$, the position and velocity of the planet and test particles are recorded every $t_{dump}$. At the end of each simulation, each $t_{dump}$ frame is stacked to obtain a final map of the dust distribution as if test particles were constantly released from the parent body belt every data dump, $t_{dump}$. This method allows us to mimic the continuous creation of small grains resulting from planetesimal collisions inside the parent body belt.

In the simulations, we initially assume that 2000 parent bodies are uniformly distributed in a region of width $\Delta r$ determined by the disc parameters of ~\cite{2010AJ....140.1051K,2012AJ....144...45K} (see Table~\ref{Table1}) with a maximal inclination of $1^{\circ}$. Although the final results are independent of this initial choice \citep{2015arXiv150908589T}, we assume the initial parent body belt of HD~202628 to be dynamically warm ($0<e<0.3$) expected from a system hosting a massive giant planet. However since HD~207129 is suspected to host a smaller planet, we assume the parent body belt to be dynamically cold with quasi circular orbits ($0<e<0.04$). Through a collisional cascade, we assume that each parent body releases a massless test particle, thus simulating the small grains traced by scattered light. Initially, the grains have identical position and velocity to their parent bodies, but because the massless test particles are sensitive to radiation forces, they will evolve on different orbits to that of their parent bodies.

The $\beta$ parameter is determined using Equation (1) with the minimal grain size and the stellar parameters obtained by ~\cite{2010AJ....140.1051K,2012AJ....144...45K}. By fitting simultaneously the scattered light image and the spectral energy distribution, their model indicates a minimum grain size of 2.8 $\mu$m. Assuming a density of $\rho=2.5$~g/cm$^{3}$ for a silicate grain and a total luminosity, $L_{\ast}$, given by $L_{\ast}/$L$_{\odot}$=($M_{\ast}/$M$_{\odot}$)$^{3.5}$ and using the stellar parameters from Table~\ref{Table1}, we find $\beta$ values for both HD~202628 and HD~207129 between $\sim$0.08--0.09.

We run each simulation for a duration corresponding to $20~t_{sec}$ where $t_{sec}$ is the secular timescale of the planet, and  with a timestep, $\Delta t$, taken to be a $30^{th}$ of the planet's period, $P_{p}$ -- see Appendix A for a discussion on our choice of  $\Delta t$ and simulation duration. During the simulation, we chose to record the test particle positions after $t_{init}=0$ (meaning that particles and planet position are recorded from the beginning of the simulation) and then save the positions every $t_{dump}=4/3~P_{p}$ -- see Appendix B for a discussion on our choice of $t_{init}$ and $t_{dump}$.

\subsection{Scattered light modeling}
To compare our simulation results with the observations of HD~202628 and HD~207129, we use the 3D Monte Carlo radiative transfer code \textit{MCFOST}~\citep{2006A&A...459..797P} to produce synthetic images.

While our simulations use massless test particles to represent dust grains,the radiative transfer code requires to know the number density of the particle distribution in order to estimate the scattered light emission. Therefore we assign a mass to each test particles recorded in the simulation to convert our simulated distribution maps into density maps using the following assumption: (i) the number of particles binned on the distribution map from the simulations represents the total mass of the disc, and (ii) the total mass of dust of about one lunar mass for both discs, as derived for HD~207129 from the infrared images by ~\cite{2012A&A...537A.110L}. The 3D density structure that results from our dynamical simulations is then mapped onto a cylindrical grid and read by \textit{MCFOST}. Considering radiation from the stellar photosphere and the dust, \textit{MCFOST} derives the temperature and radiative structure of the disc assuming spherical homogeneous grains. Via a ray-tracing method to follow the path of stellar photons moving throughout the disc, synthetic images can then be obtained. We set the stellar properties and disc orientation (inclination and position angle) from observations - see Table~\ref{Table1} - to create the synthetic images.

The dimensions of the grid used by \textit{MCFOST} sets the resolution of the synthetic images. For our 3D grid, we chose 90 radial bins between $r =120-280$ AU for HD~202628 and $r =110-225$ AU for HD~207129, producing an average radial bin size of $\sim$ 1.5 AU, which is close to the resolution of the \textit{HST} images for these two objects. We use 120 azimuthal bins between 0 to 2$\pi$, and 60 vertical bins between $z \pm 20$~AU. The size of the image is set to $550 \times 550$~AU to cover the disc surface in size, and we create synthetic images $512 \times 512$ pixels, with a resolution of 1.1 AU per pixel, again close to the resolution of \textit{HST}.

\subsection{Image fitting method}
To compare the synthetic resulting image from each simulation with observations, we determine the three following parameters for each model: the peak brightness, $r_{0}$, the disc width ratio, $\Delta r/r_{0}$, and offset of the disc, $\delta$. We then estimate the best fit model using a $\chi^{2}$ method and a Bayesian probability distribution.

To determine the radial location of the disc peak brightness, $r_{0}$, and width, $\Delta r$, we use the following method. First the synthetic image is radially cut into 90 bins or rings (matching the radial bins of the cylindrical grid of \textit{MCFOST}), and the surface brightness in each ring is then azimuthally averaged. To increase the quality of the fit, we use a spline method to interpolate between the 90 bins to obtain a set of 500 radial bins in $r$ with corresponding averaged surface brightness $F(r)$. The peak brightness location, $r_{0}$, corresponds to the radial bin where the maximal surface brightness, $F_{0}$, is measured, while the width, $\Delta r$ of the disc is defined as the FWHM measured from the surface brightness profile.

To determine the offset of the disc, we first divide the disc image azimuthally into 120 bins (matching the azimuthal bins of the cylindrical grid of \textit{MCFOST}) and then extract the peak emission from each of the 120 bins. From the coordinates of the 120 emission peaks, we use a least-squares deviation method to fit an ellipse that passes through all the emission peaks in each azimuthal bin. This fit provides four parameters: semi-minor and semi-major axes, $b$ and $a$, and the coordinates of the center of the ellipse offset compared to the location of the star, ($x_{\rm off},y_{\rm off}$), from which we can derive the disc offset, $\delta=\sqrt{x_{\rm off}^2+y_{\rm off}^2}$.

\subsubsection{$\chi^{2}$ Method: }
Once the three disc parameters have been derived from the synthetic images, we compute the $\chi^{2}_{X}$ for each parameter given by:
\begin{equation}
\chi^{2}_{X}=\frac{(X_{Obs}-X_{Sim})^{2}}{\sigma_{X/Obs}^{2}+\sigma_{X/Sim}^{2}}
\end{equation}
for each of the three parameters $X=r_{0}$, $\Delta r/r_{0}$ and $\delta$, the subscripts $Obs$, $Sim$ are respectively the parameter value from the scattered light observation and the simulation, and $\sigma_{X/Obs}$ and $\sigma_{X/Sim}$ are respectively the standard deviation of the parameter from the scattered light observation and the simulation.

To estimate the simulation standard deviation, $\sigma_{X/Sim}$ of each $X$ parameter, we use a bootstrap method which consists of randomly selecting and replacing some values to resample a set of data points, allowing us to statistically estimate the accuracy of fitting on the original set of data points. For the disc width ratio and peak brightness location, we bootstrap and resample our 500 data points of surface brightness as a function of radius, $F(r)$, to create an additional 5000 samples, and we then extract the peak location and disc width ratio for each sample. The standard deviation of the two parameters across the 5000 samples gives $\sigma_{r_{0}/Sim}$ and $\sigma_{(\Delta r/r_{0})/Sim}$. In the case of the disc offset, we bootstrap and resample our 120 datapoints corresponding to the peak emission coordinates in each of 120 azimuthal bins to create an additional 5000 samples. We then fit an ellipse via a $\chi^{2}$ to obtain the value of offset for each sample. The standard deviation of the offset across the 5000 samples gives $\sigma_{\delta/Sim}$.

Unfortunately, there is little information in ~\cite{2010AJ....140.1051K,2012AJ....144...45K} regarding uncertainties in the disc parameters derived from the scattered light observations, with the exception of the disc eccentricity of HD~202628. Between $120< r< 280$~AU, the observed brightness profile of HD~202628 from \cite{2012AJ....144...45K} consists of 30 datapoints (see their Figure 5), while our disc has three times more datapoints with 90 radial bins between $120< r< 280~$AU. Therefore we assume that $\sigma_{r_{0}/Obs}=\sqrt{3}\sigma_{r_{0}/Sim}$ and $\sigma_{(\Delta r/r_{0})/Obs}=\sqrt{3}\sigma_{(\Delta r/r_{0})/Sim}$. The eccentricity uncertainty ($\pm 0.02$) estimated in Krist et al. is similar to the value we derived from the standard deviation of the offset and the peak location of our bootstrap sample, and we therefore assume that  $\sigma_{\delta/Obs}=\sigma_{\delta/Sim}$.

We derive the total $\chi^{2}_{tot}$ of the model given by the sum of $\chi^{2}_{X}$ of each parameters:
\begin{equation}
\chi^{2}_{tot}=\chi^{2}_{\delta}+\chi^{2}_{r_{0}}+\chi^{2}_{\Delta r/r_{0}}
\end{equation}
The best fit model corresponds to the model with the minimum $\chi^{2}_{tot}$.

\subsubsection{Bayesian probability distribution}
To ensure that the range of initial planetary parameters covers a broad enough portion of the parameter space and to therefore assess the quality of our best fit model \citep{2007A&A...469..963P}, we can check the relative probability of occurrence of each grid point in the parameter space using a Bayesian analysis \citep{1997ApJ...489..917L}. 

A Bayesian analysis requires an assumption of the prior distribution for the planetary parameters. Using the first set of simulations with the Rodigas et al. predicted planetary parameters as initial conditions as our only preliminary information, we use an uniform distribution for the semi-major axis and eccentricity, corresponding to uniform sampling of the parameter space, and a non-uniform distribution for the planetary mass when required. For both discs, the semi-major axis uniform distribution uses the minimal predicted semi-major axis as the minimal boundary. The upper boundary is empirically defined as the semi-major axis for which the planet crosses the inner edge of the disc. Given that the predicted eccentricity for HD~207129 is strictly an upper limit, we use the predicted value as the upper boundary for the eccentricity uniform distribution and set the minimal boundary to be the lowest possible value. For HD~202628, we use the predicted eccentricity minus  $1\sigma$ ($ \pm 0.02$ as estimated by Krist et al.) as the minimal boundary of the distribution, and chose a reasonable upper value at $4\sigma$ from the predicted value. In addition, for HD~202628 we explore a non-uniform distribution for the planetary mass as we use an irregular spaced grid to decrease the total number of models explored and to quickly converge towards the best fit value. The exact distributions for the planetary parameters are defined in Section 5.2.1 for HD~202628 and 5.2.2 for HD~207129.

The probability of a specific grid point is given by $P \sim e^{-\chi^{2}/2}$. In our study, the parameter space is either a 2D grid or 3D grid with various initial planetary semi-major axis, $a_{p}$, planetary eccentricity, $e_{p}$, and planetary mass, $m_{p}$, values which are input into the simulations. Each $\left( a_{p},e_{p},m_{p} \right)$ triplet forms a different model for which a $\chi^{2}_{tot}$ value is derived. The 3D probability is therefore $P_{3D}(a_{p},e_{p},m_{p}) \sim e^{-\chi^{2}/2}$, and after normalization, we can marginalize each parameter to obtain its probability of occurrence: 
\begin{eqnarray}
P_{1D}(e_{p}) & \sim & \sum\limits_{a_{p},m_{p}} P_{3D}(a_{p},e_{p},m_{p}), \\
P_{1D}(a_{p}) & \sim & \sum\limits_{e_{p},m_{p}} P_{3D}(a_{p},e_{p},m_{p}) \\
P_{1D}(m_{p}) & \sim & \sum\limits_{a_{p},e_{p}} P_{3D}(a_{p},e_{p},m_{p}).
\end {eqnarray}
We can therefore check that (i) the best fit model corresponds to parameters for which the probability of occurrence is maximal, and (ii) evaluate the quality of the fit depending on how this probability distribution peaks around the maximal value.

\section{Results}
\subsection{Testing Rodigas et al. predictions}
 For both debris discs, we start by running a simulation using the observed width, $\Delta r$ and eccentricity, $e$,  of the disc from Table~\ref{Table1} to set the initial semi-major axis and eccentricity for the test particles. We use the parameters predicted by ~\cite{2014ApJ...780...65R} as initial conditions for the planet parameters $m_{p}$, $a_{p}$ and $e_{p}$ -- see Table~\ref{Table2}. We run the simulations using the simulation parameters described in Section 4.2, before creating a synthetic image with \textit{MCFOST} and finally, fitting the image parameters according the method described in Section 4.4 to derive the $\chi^{2}$ of the model. 

\begin{figure}
\begin{center}
  \includegraphics[width=90mm,height=67mm]{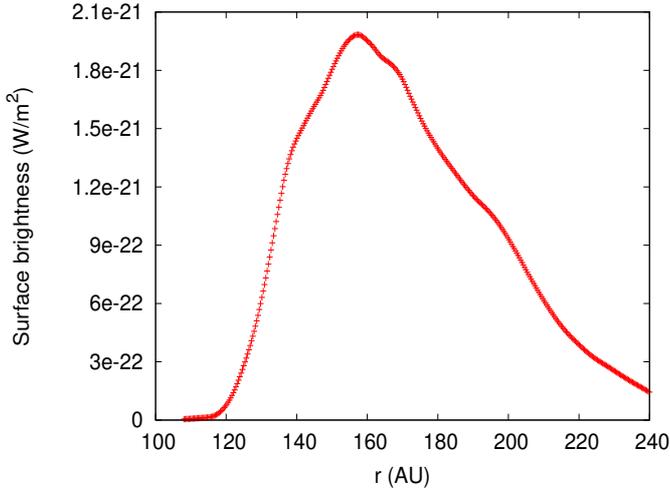}                            
  \caption{Deprojected and azimuthally averaged radial surface brightness profile (in W/m$^{2}$) of the standard model of HD~202628 using the initial planetary parameters from Rodigas et al. (2014).} 
\label{fig:1}  
\end{center}
\end{figure}

In Figure~\ref{fig:1}, we present the results of the standard model for HD~202628 with the initial planet parameters $m_{p}=15.4$~M$_{\rm J}$, $a_{p}=71$~AU and $e_{p}=0.18$ given by ~\cite{2014ApJ...780...65R}. In order to compare with observations of \cite{2012AJ....144...45K}, we first fit an ellipse on the inner edge of the belt on the deprojected image, which corresponds to the inner radius of the disc with a surface brightness equal to $50\%$ of maximum peak brightness. We then  measure the offset, $\delta$, of the ring constituting the inner edge, as well as derive the peak location of emission, $r_{0}$, and disc FWHM, $\Delta r$, from the deprojected and azimuthally averaged surface brightness profile. We then estimate the eccentricity, $e=\delta/r_{0}$. We found that the disc can be described by an ellipse with $e$ = 0.08 and $\delta= 11.8$ AU. The brightness profile peaks at $r_{0}=$157.5 AU with $\Delta r=60$~AU, and the corresponding $\Delta r/r_{0}$ ratio is 0.41. The resulting $\chi^{2}$ of the standard model is $\chi^{2}=2088$. We conclude that the resulting structure of the HD~202628 system proposed by the \cite{2014ApJ...780...65R} planet predictions produces a belt with the expected width -- see Table~\ref{Table1}, but with the mean radius, $r_{0},$ 20 AU closer to the star than observed. This planetary configuration also results in a smaller disc eccentricity and smaller offset than observed -- see Table~\ref{Table3} for a summary.

\begin{table*}
\renewcommand{\arraystretch}{1.0}
\caption{Comparison between the observed disc parameters in scattered light and the disc parameters resulting from simulations when testing the planetary parameters predicted by Rodigas et al. (2014).}
\label{Table3}
\centering
\begin{tabular}{cccccccccc}
\hline
Star & $e_{Obs}$ & $e_{Sim}$ & $\delta_{Obs}$ (AU) & $\delta_{Sim}$ (AU) & $r_{0, Obs}$ (AU) & $r_{0, Sim}$ (AU) & $\Delta r/r_{0, Obs}$ & $\Delta r/r_{0, Sim}$ & $\chi^{2}$\\
\hline
HD~202628 & 0.18 & 0.08 & 28.4 & 11.8 & 182 & 157.5 & 0.4 & 0.41 & 2088\\
HD~207129 & $<$0.08 & 0.07 & $<$13.5 & 11.0 & 163 & 156.0 & 0.2 & 0.18 & 10 \\
\hline
\end{tabular}
\end{table*}

\begin{figure}
\begin{center}
\includegraphics[width=90mm,height=67mm]{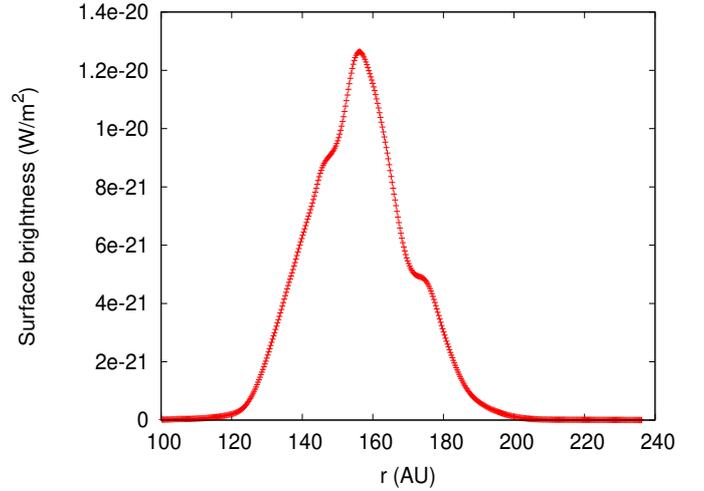}           
  \caption{Same than Figure 1 for the standard model of HD~207129.} 
\label{fig:2}  
\end{center}
\end{figure}

In Figure~\ref{fig:2}, we present the results of the standard model HD~207129 with the initial planet parameters $m_{p}=4.2$~M$_{\rm J}$, $a_{p}=92$~AU and $e_{p}=0.08$ given by ~\cite{2014ApJ...780...65R}. In order to compare with observations of ~\cite{2010AJ....140.1051K}, we follow the same method as for HD~202628, except that the ellipse is now fitting the center of the emission belt the disc (following Krist et al. method). We found that the bright emission belt of the disc can be described by an ellipse at $r_{0}=156$~AU with $\Delta r= 28$~AU, $e$ = 0.07 and $\delta =11$~AU. The corresponding $\Delta r/r_{0}$ ratio is 0.18. The resulting $\chi^{2}$ of the standard model is  $\chi^{2}=10$. We conclude that the structure of the HD~207129 system proposed by ~\cite{2014ApJ...780...65R} provides a good match in terms of width, eccentricity and offset (although the values are very close to the upper limits), but also produces a belt with a emission peak 7 AU closer to the star than observed -- see Table~\ref{Table3} for a summary.

\subsection{Finding the best fit}
To try to improve the model fit to the observations, additional simulations were run using a broader range of planetary parameters depending of the outcome of the Rodigas et al. standard models. If, for example, the resulting mean radius of the disc is located closer from the star than observed, then either the planetary semi-major axis and mass, $a_{p}$ and $m_{p}$, are increased to strengthen the grains gravitational scattering by the planet. If the impact of the planet on the disc structure was too weak, producing a smaller eccentricity and offset than observed, it is likely that the planet is either too far from the disc (since the timescale of secular perturbation is inversely proportional to the location of the planet) or has too low an orbital eccentricity. We therefore explored a range of possible semi-major axes, $a_{p}$, starting from the minimum value provided by ~\cite{2014ApJ...780...65R}, and also explored a range of possible planet eccentricities, $e_{p}$, as well as planetary masses, $m_{p}$. For both systems we needed to run additional simulations to better match the observations, and a total of 190 runs of simulations were performed -- see Table~\ref{Table5} and Table~\ref{Table6} for a summary.

\subsubsection{HD~202628}
The standard model for HD~202628 resulted in a disc with a peak emission too close from the star and an eccentricity and offset smaller than observed, suggesting that the planet was located too close to the star with an eccentricity is slightly too small. We therefore conducted a set of simulations with the planetary eccentricity, $e_{p}$, ranging from 0.16 to 0.26 in steps of 0.02, and semi-major axis, $a_{p}$, ranging from 71 AU to 111 AU in steps of 10 AU. Simulations with $a_{p} > 120$~AU (for which the planet becomes disc crossing) led to a continuous disc with a decreasing brightness profile, which clearly do not match the observed ring with its sharp inner edge and bell-shaped brightness profile. Therefore an upper limit of $a_{p}=111$~AU must be set. As no satisfactory match could be found using the predicted planetary mass from ~\cite{2014ApJ...780...65R}, we additionally explored a range of planetary mass: 15.4, 9.9, 4, 3 and 2~M$_{\rm J}$. Simulations with $m_{p} < 2$~M$_{\rm J}$ led to a peak emission too close from the star for any value of $a_{p}$ within the range of $71-111$~AU, and therefore $2$~M$_{\rm J}$ was set as the minimal planetary mass. This lead to a total of 150 simulations -- see Table~\ref{Table5}.

\begin{figure*}
\begin{center}
  \subfloat{\includegraphics[width=90mm,height=67mm]{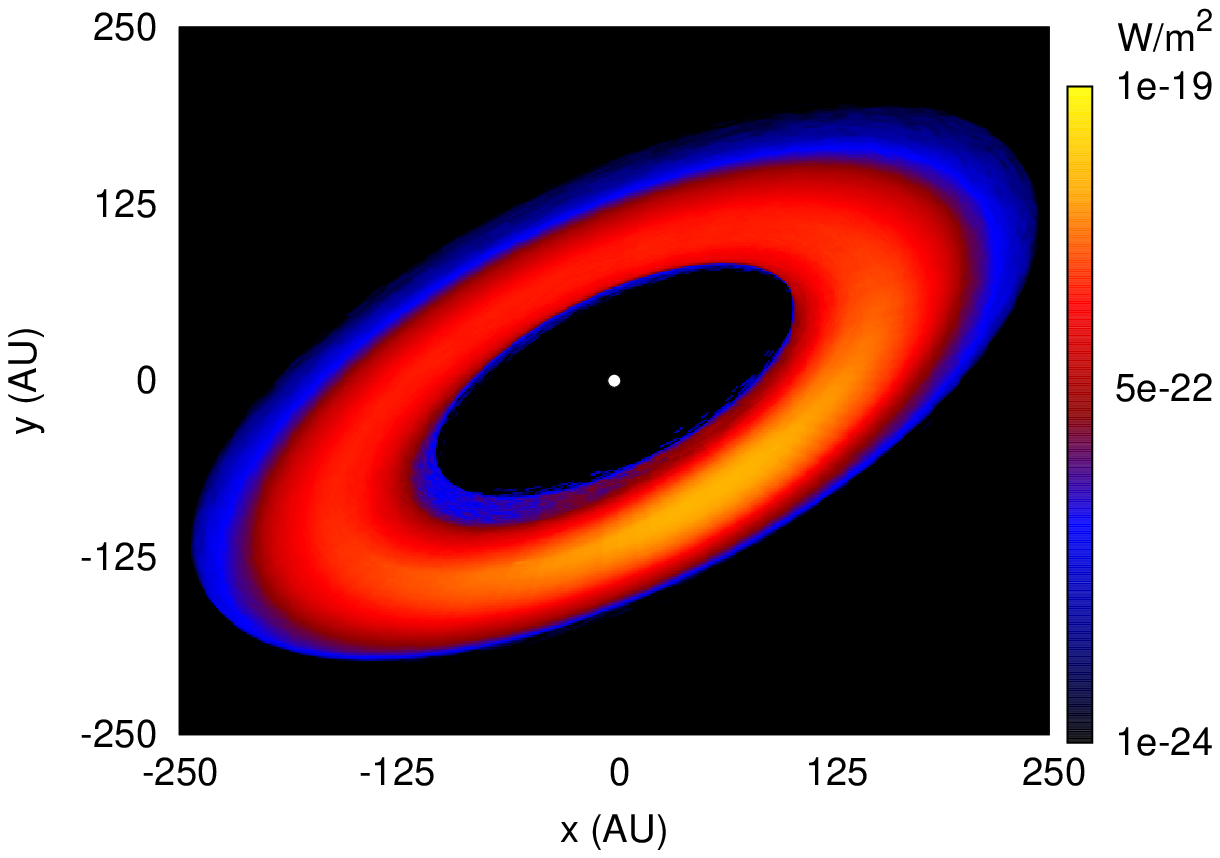}}                
  \subfloat{\includegraphics[width=90mm,height=67mm]{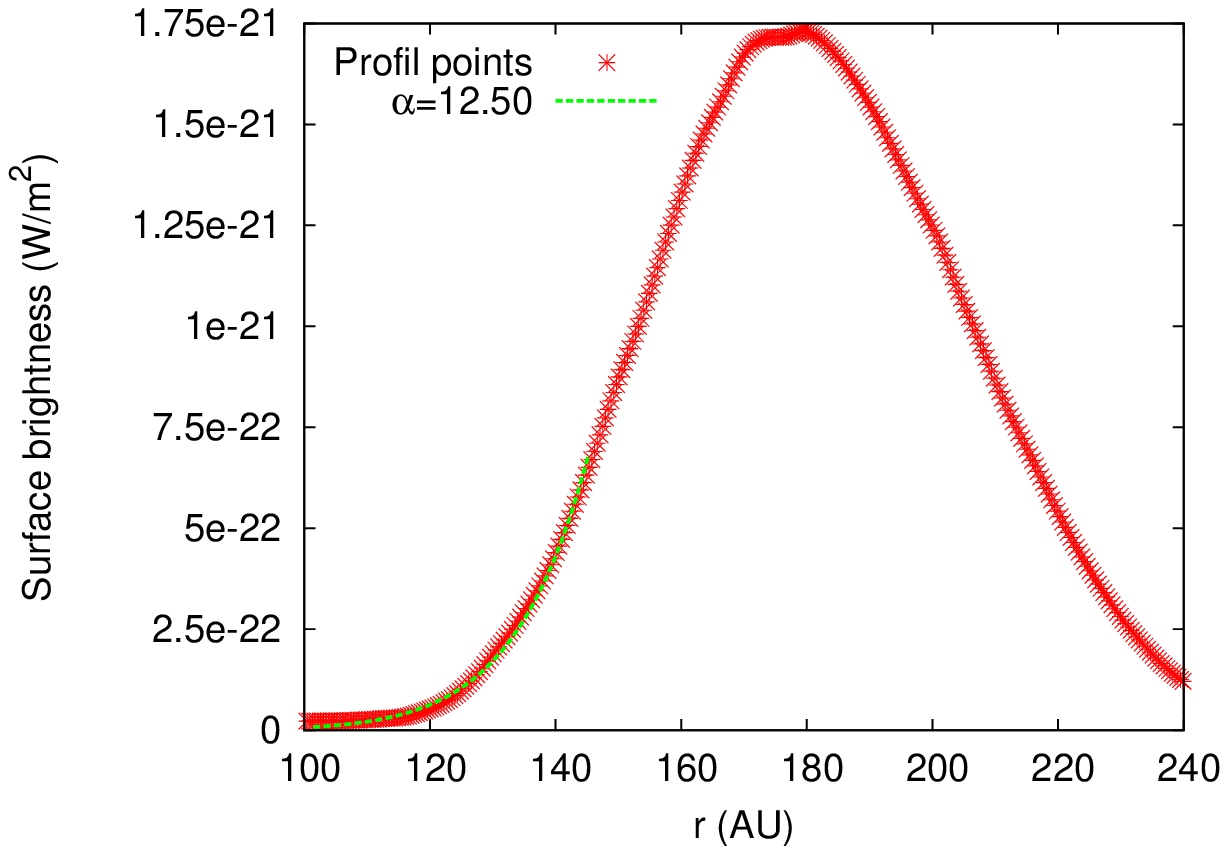}} \\           
  \caption{Result for the best-fit model HD~202628-21: (left) synthetic image at 1.03 $\mu$m roughly aligned with the image from Krist et al., (right) deprojected and azimuthally averaged radial surface brightness profile. The green line is to the power law fit to the inner edge.} 
\label{fig:3}  
\end{center}
\end{figure*}

\begin{figure*}
\begin{center}
  \subfloat{\includegraphics[width=60mm,height=45mm]{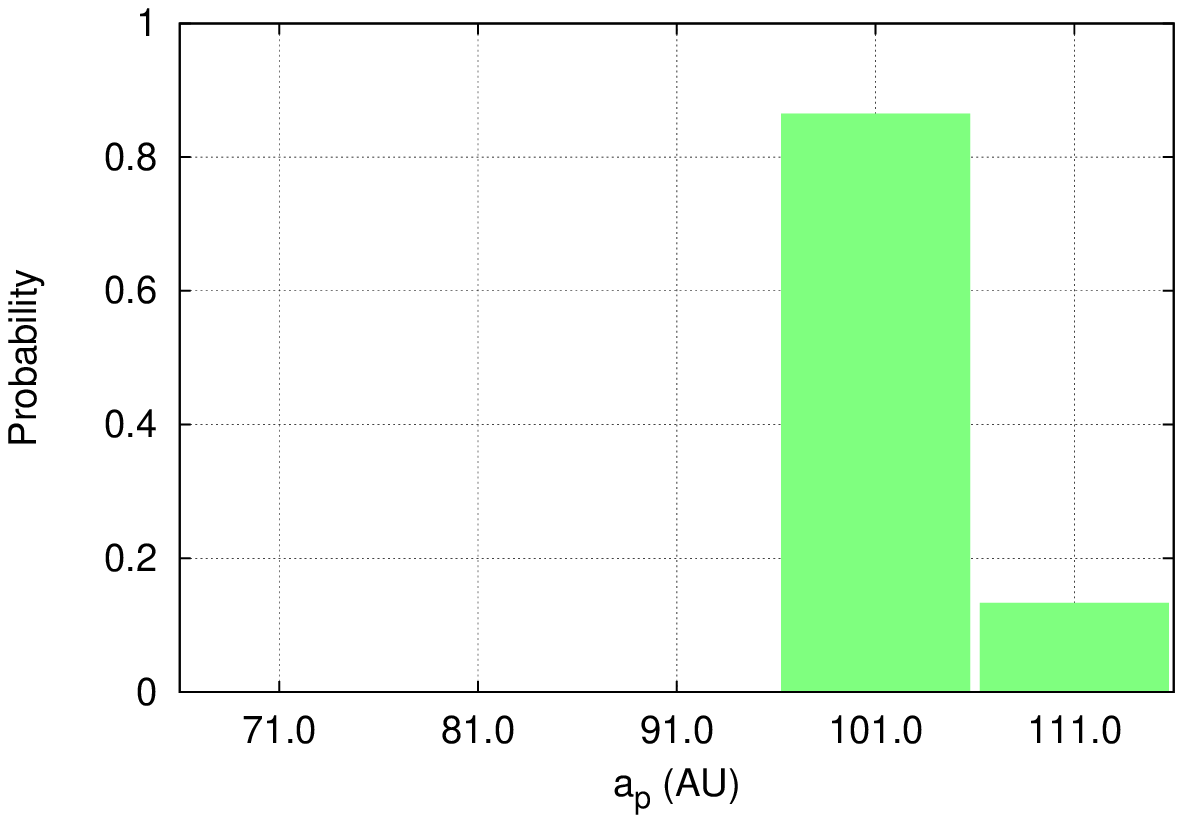}}                
  \subfloat{\includegraphics[width=60mm,height=45mm]{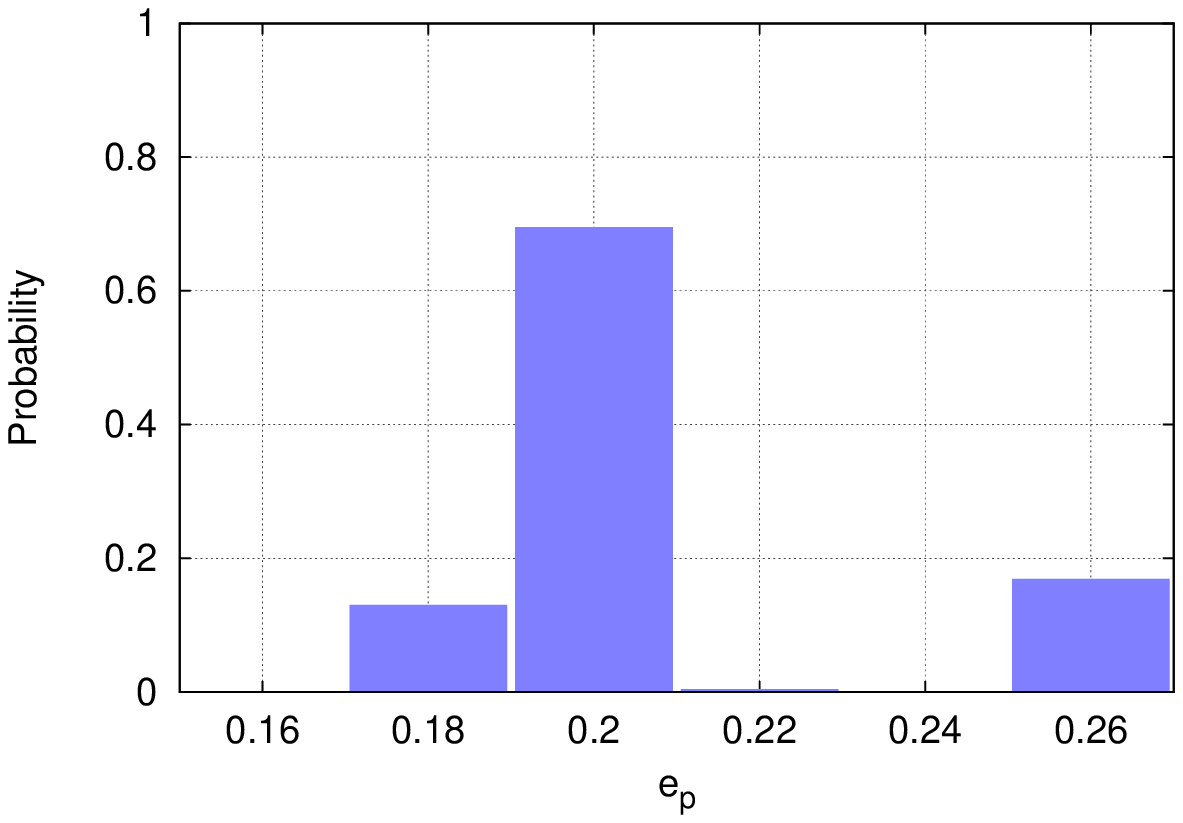}}
  \subfloat{\includegraphics[width=60mm,height=45mm]{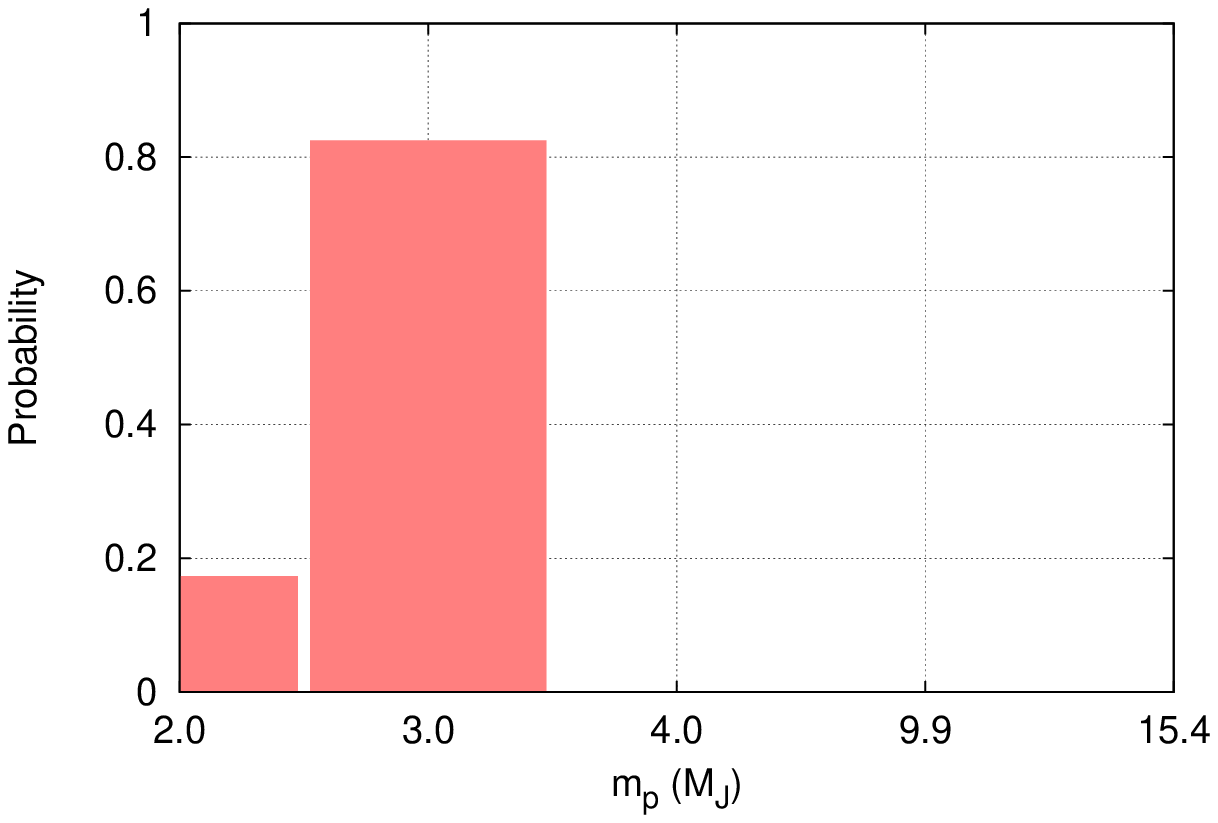}}
 \\
  \subfloat{\includegraphics[width=60mm,height=45mm]{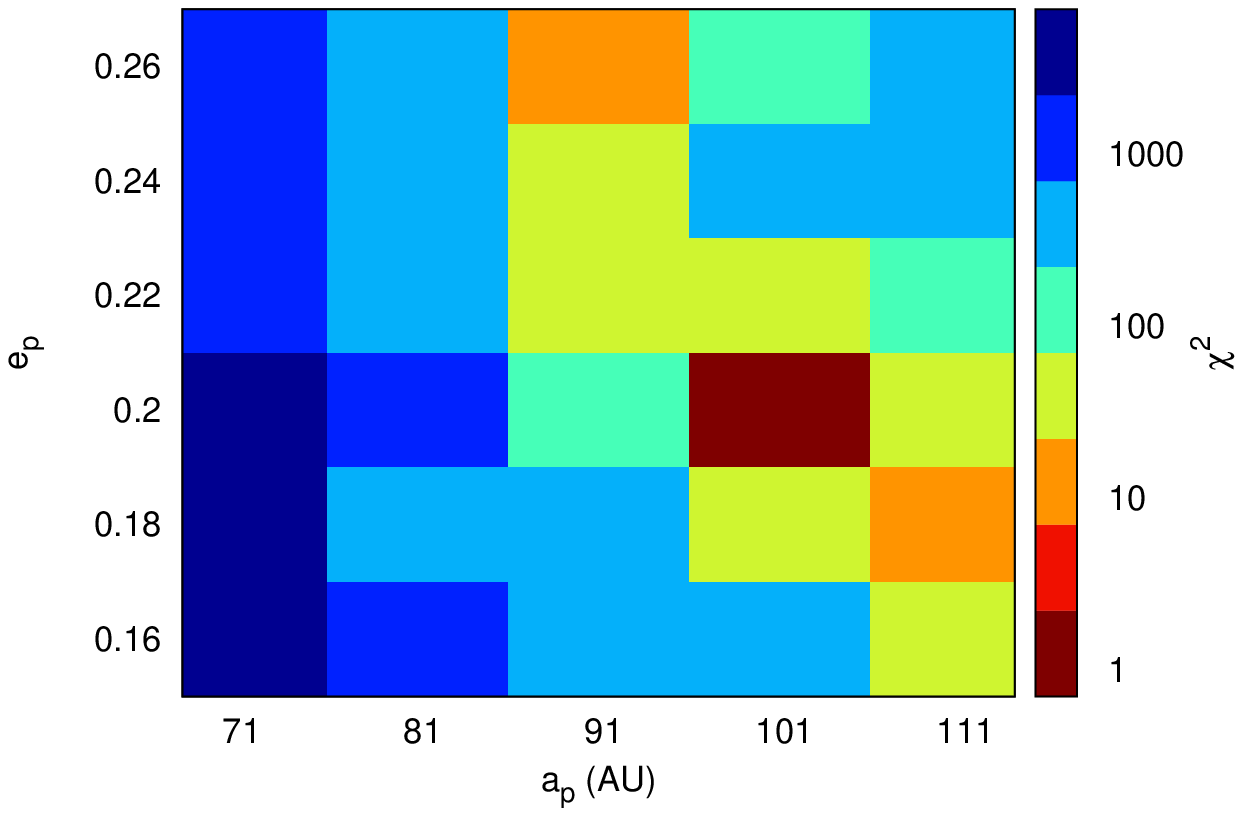}}                
  \caption{(top) Bayesian probability distribution from the set of 150 simulations of HD~202628 for (left) the parameter $a_{p}$, (center) for the parameter $e_{p}$ and $m_{p}$ (right). (bottom) $\chi^{2}$ map of the parameter space explored in the 30 simulations with $m_{p}=3$~M$_{\rm J}$, the colorbar indicates the $\chi^{2}$ value for each model.} 
\label{fig:4}  
\end{center}
\end{figure*}

Our best fit between the observations and simulations was the model HD202628-21 ($\chi^{2}=0.78$): a 3~M$_{\rm J}$ planet located at $a_{p}=101$~AU with eccentricity $e_{p}=0.20$ which resulted in a disc with an offset of $\delta =28.5$~AU, eccentricity $e=0.16$, peak location $r_{0}=$179.3~AU and width, $\Delta_{r}=60$~AU. The corresponding disc width ratio, $\Delta r/r_{0}$, of the disc is 0.34. We note that the planetary mass is 5 times smaller than the predicted maximal mass by ~\cite{2014ApJ...780...65R}. We fit a radial power-law to the inner edge of the brightness profile of the disc in $r^{\alpha}$ with $\alpha=12.5$, which is still consistent with the value of $\alpha=12$ derived by \cite{2012AJ....144...45K}. The resulting \textit{MCFOST} synthetic image and the surface brightness profile is given in Figure \ref{fig:3}. The $\chi^{2}$ map for the set of 30 simulations with $m_{p}=3$~M$_{\rm J}$ is given in Figure~\ref{fig:4}, along with the Bayesian profiles of $a_{p}, e_{p}$ and $m_{p}$ marginalized over the 150 simulations. The Bayesian distribution peaks for the best fit value of $a_{p}=101$~AU, $m_{p}=3$~M$_{\rm J}$ and $e_{p}=0.20$.

\subsubsection{HD~207129}
The standard Rodigas et al. model resulted in a disc with a peak emission too close to the star and an eccentricity and offset close to the predicted upper limit, suggesting that the planet in the standard model is located too close to the star and has too high an eccentricity. In order to find a better fit, we ran 40 simulations with planetary eccentricity, $e_{p}$, ranging from 0.08 to 0.01 in steps of 0.01, and a semi-major axis, $a_{p}$, ranging from 92 AU to 107 AU by step of 5 AU.

\begin{figure*}
\begin{center}
  \subfloat{\includegraphics[width=60mm,height=45mm]{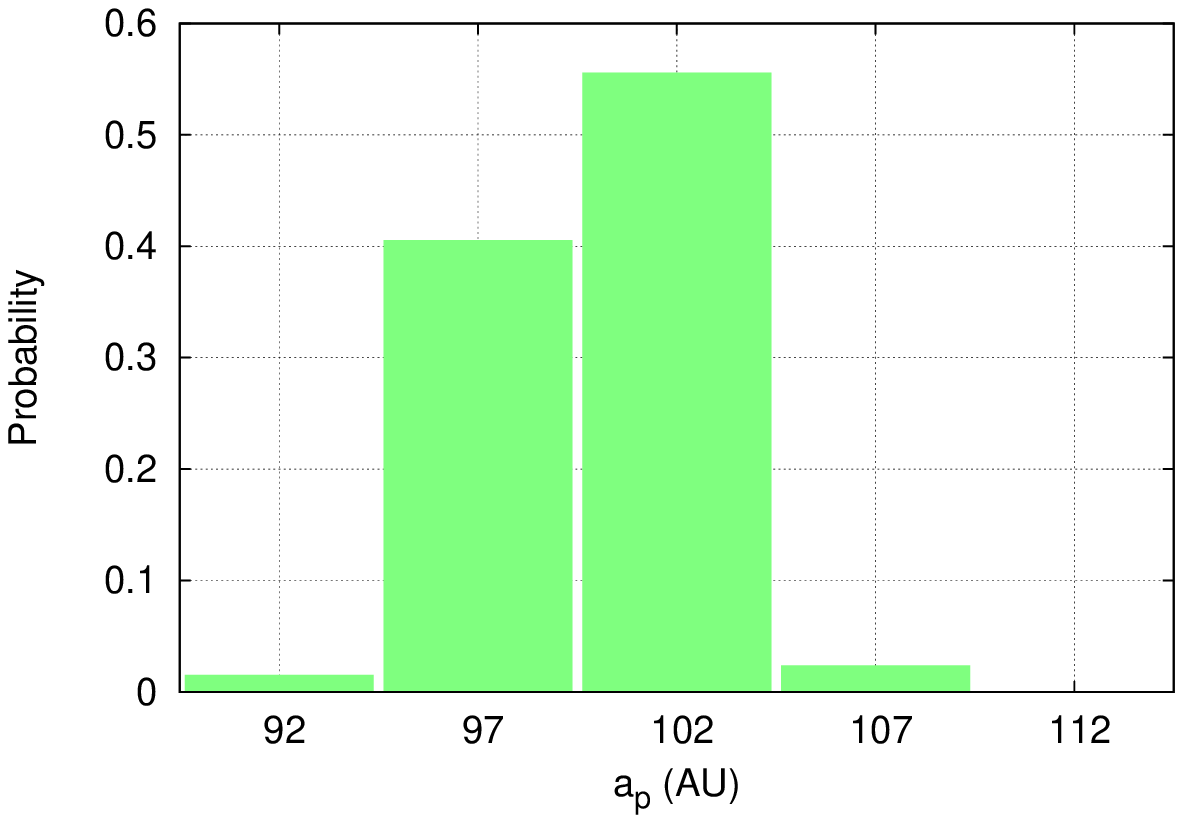}}                
  \subfloat{\includegraphics[width=60mm,height=45mm]{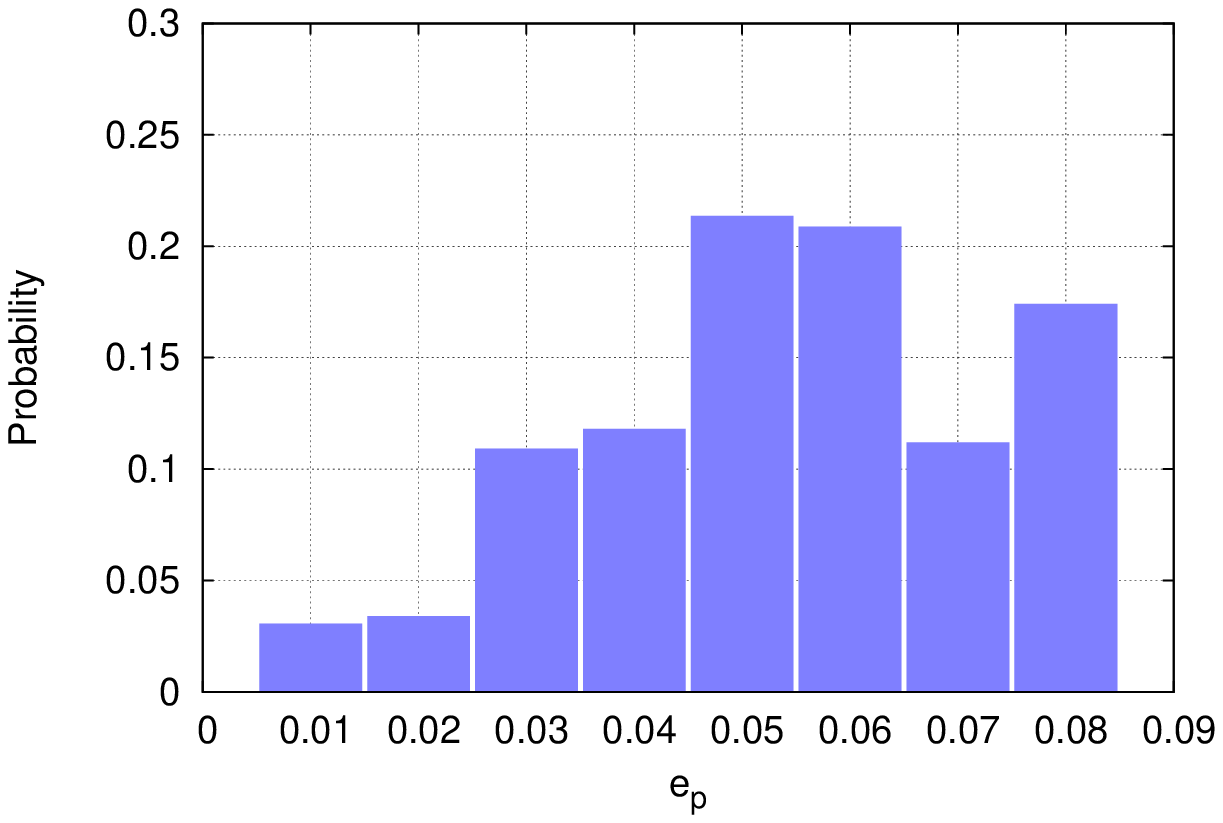}} 
  \subfloat{\includegraphics[width=60mm,height=45mm]{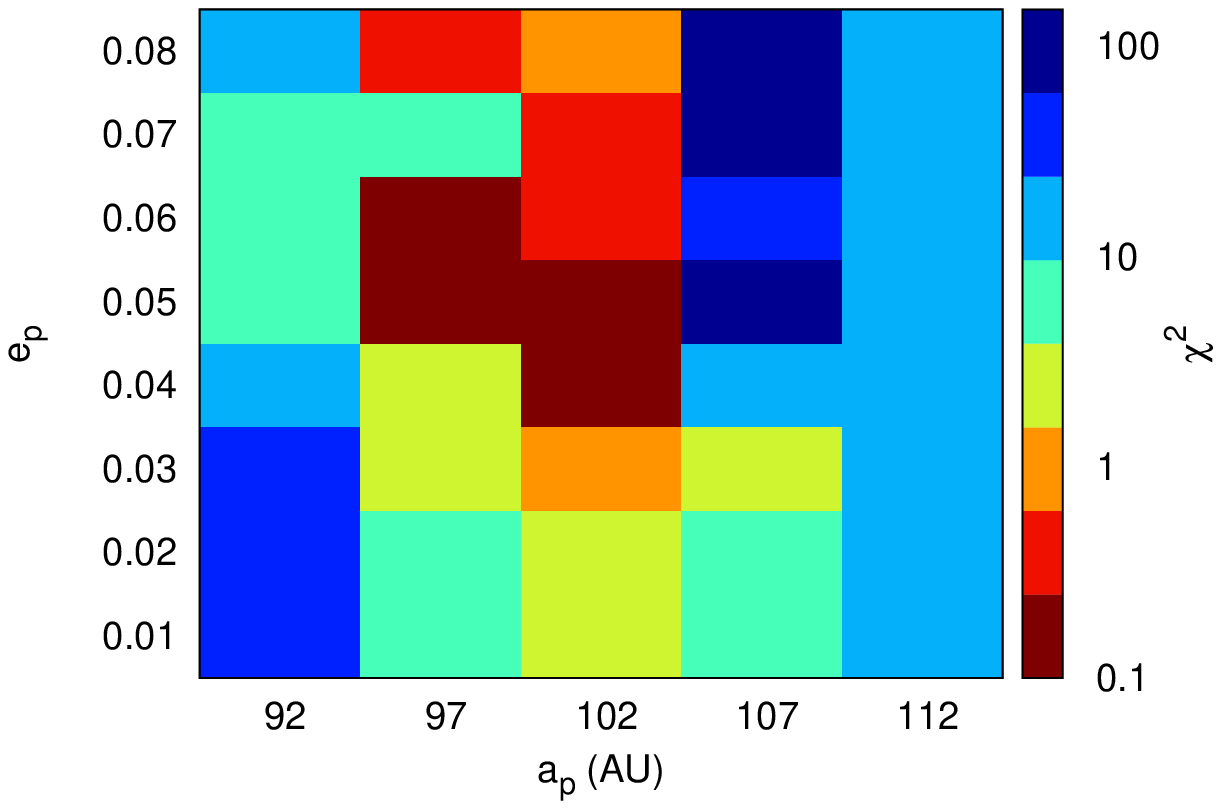}}                
  \caption{Bayesian probability distribution from the set of 40 simulations of HD~207129 for the parameters (left) $a_{p}$, (center) $e_{p}$ with a 4.2 M$_{\rm J}$ planet. (right) $\chi^{2}$ map of the parameter space explored in the 40 simulations, the colorbar indicates the $\chi^{2}$ value for each model.} 
\label{fig:5}  
\end{center}
\end{figure*}

\begin{figure*}
\begin{center}
  \subfloat{\includegraphics[width=90mm,height=67mm]{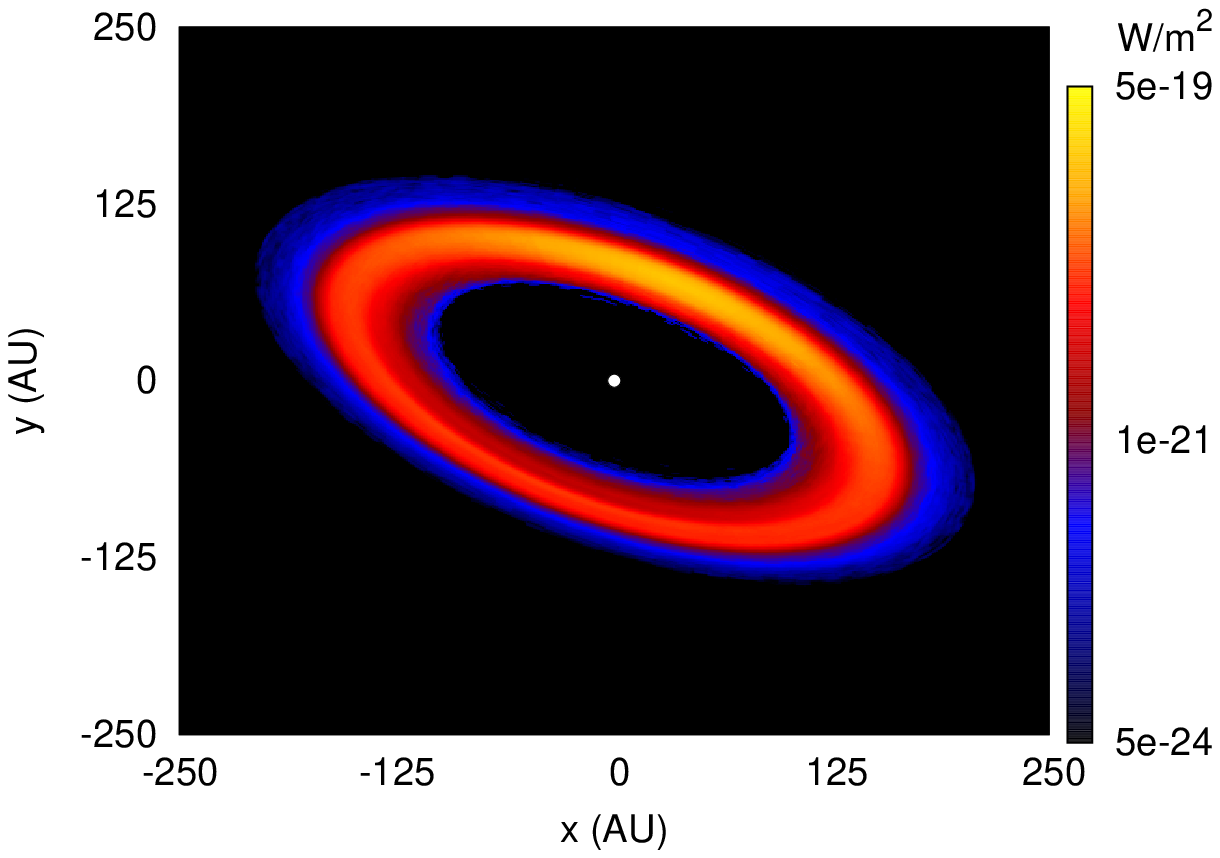}}                
  \subfloat{\includegraphics[width=90mm,height=67mm]{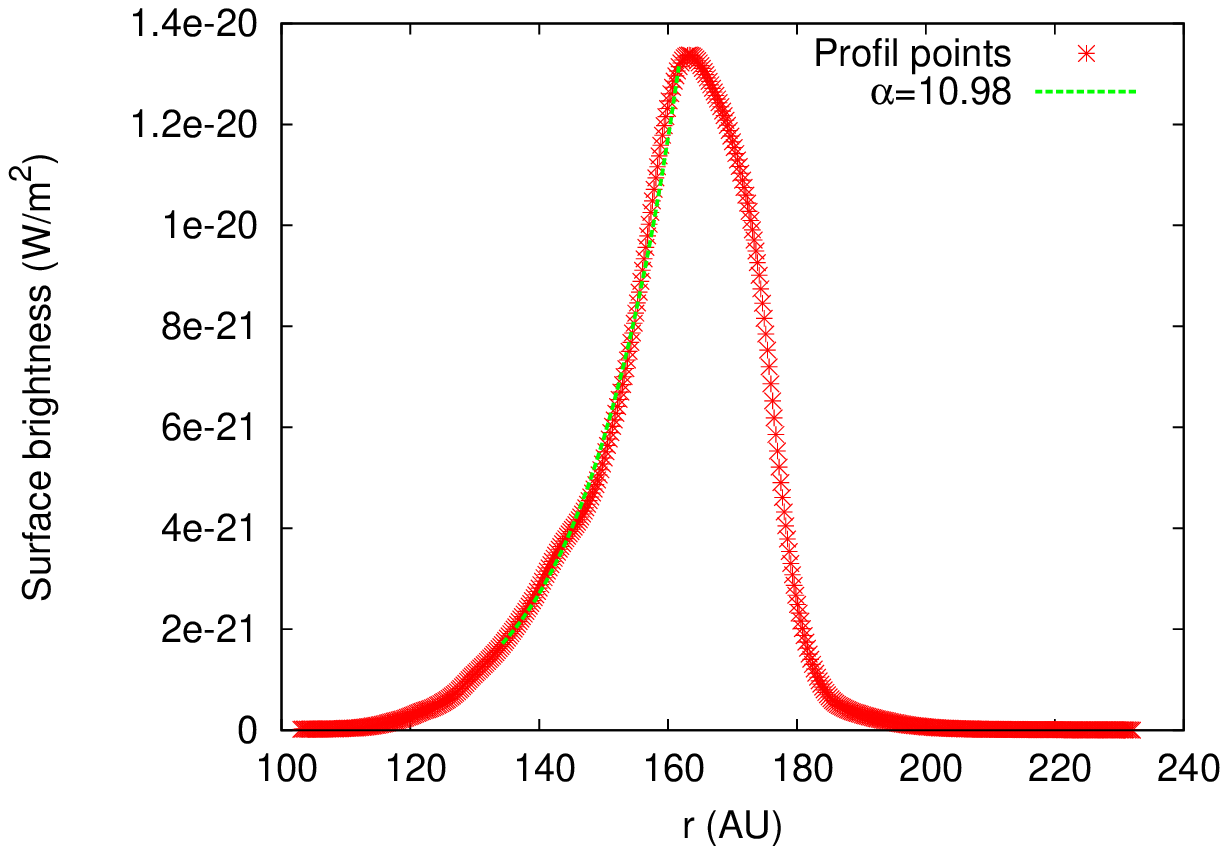}} \\           
  \caption{Result for the best-fit model HD~207129-20: (left) synthetic image at 1.03 $\mu$m roughly aligned with the image from Krist et al., (right) deprojected and azimuthally averaged radial surface brightness profile. The green line is to the power law fit to the inner edge. } 
\label{fig:6}  
\end{center}
\end{figure*}

We found good agreement between the observations and models 11, 12, 20 and 21 ($0.07<\chi^{2}<0.2$): a planet located at $a_{p}=97$~AU with $e_{p}=0.05-0.06$  resulted in a disc with an offset $\delta \sim 10$~AU, an eccentricity $e$ of 0.06, with a peak location $r_{0} \sim$163~AU and a $\Delta_{r}=27$~AU ($\Delta r/r_{0} \sim 0.17$), while a planet located at $a_{p}=102$~AU with $e_{p}=0.04-0.05$ led to $\delta=7$~AU, $e=0.03$, $r_{0}\sim 163$~AU and $\Delta r/r_{0} \sim 0.16$ -- see Table~\ref{Table6}. We thus based our determination of the best-fit model on the Bayesian distributions marginalized over the 40 simulations in Figure~\ref{fig:5}: the distributions peak for $a_{p}=102$~AU and $e_{p}=0.05$, and we therefore use model 20 as the best fit ($\chi^{2}=0.16$), and the $\chi^{2}$ map is given in Figure~\ref{fig:5}.  Our radial power-law fit to the inner edge of the brightness profile of the disc was in $r^{\alpha}$ with $\alpha=10.98$, which is consistent with the inner edge suspected by ~\cite{2010AJ....140.1051K} when noticing the steep rise at 30 $\mu$m in the SED. The resulting \textit{MCFOST} synthetic image and the surface brightness profile are given in Figure~\ref{fig:6}. 

\section{Discussion}
\subsection{The best fit for HD~202628}
HD~202628 was recently observed with \textit{Herschel} by \cite{2013AAS...22114414S}, who noted a very sharp inner edge of the disc with a well constrained eccentricity of ~0.18. They suggested that a planetary companion with $a_{p} \ge$~100 AU could be sculpting the disc. Our best fit for this system placed the planet at 101~AU with $e_{p}=0.2$ and a resulting disc eccentricity of 0.16, supporting the suggestion of ~\cite{2013AAS...22114414S}. Those parameters are also consistent with the predictions of \cite{2015ApJ...798...83N} and \cite{2014MNRAS.443.2541P}. We found that if the planet was located beyond 120 AU, the proximity of the planet to the belt would produce a strong dust scattering, leading to a continuous dust disc from the star to the outer region of the system, which is not the structure observed by ~\cite{2012AJ....144...45K}. If $a_{p} <$ 100~AU, the secular perturbation does not reproduce a disc with eccentricity and offset high enough to match the observations. Therefore we suggest a configuration with a planet of $e_{p}=0.2$ and $a_{p} \sim 100$~AU represents the best fit to the scattered light observations of HD~202628.

\subsection{The best fit for HD~207129}
During the analysis of images taken by \textit{Herschel/PACS} at $70$ and $100$~$\mu$m, ~\cite{2012A&A...537A.110L} noted a brightness asymmetry of about $15\%$ between the north-west and south-east side of the disc. Using the code \textit{ACE} (Analysis of Collisional Evolution) to model the collisional activity of HD~207129, the authors derived an average eccentricity for the parent body belt of $e \sim$~0.05. 
In order to check our best fit for this system, we ran additional simulations of 2000 parent bodies each, with $\beta$ parameters corresponding to the  minimal and maximal grain size derived from the \textit{Herschel/PACS} emission ($s=8$ and 1000 $\mu$m) along with an intermediate size of $s=100$~$\mu$m. For the planet parameters, we use a simulation with (i) an initial planetary semi-major axis $a_{p}=97$~AU with $e_{p}=0.06$ (model 11), and (ii) a simulation with $a_{p}=102$~AU with $e_{p}=0.05$ from the scattered light best fit (model 20).
For each set of initial planetary parameters, the three distributions resulting from simulations with different grain sizes were binned into a single \textit{MCFOST} grid. 
Although each of the three simulations with a different grain size are run with an equal numbers of parent bodies, the final spatial distributions have their contribution weighted according the classical size distribution \citep{1969JGR....74.2531D} when binned into the grid, as expected for a steady-state system. Using this stacked distribution, \textit{MCFOST} therefore derives the temperature structure for a disc with a density distribution estimated by the three different grain sized populations. We then created synthetic images at $70$, $100$ and $160$~$\mu$m which were convolved with a gaussian beam of FWHM equal to the resolution of the \textit{Herschel/PACS} instrument.\\

\begin{figure*}
\begin{center}
  \subfloat{\includegraphics[width=90mm,height=67mm]{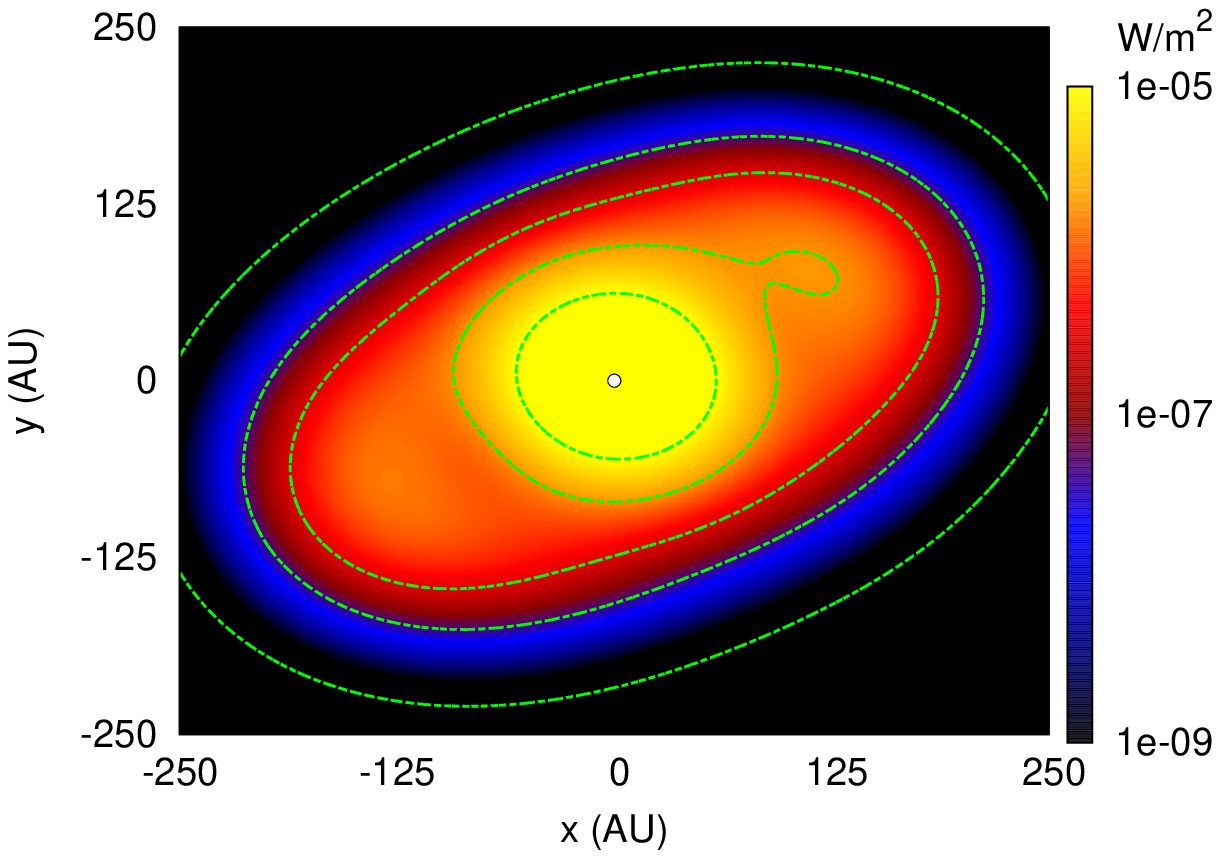}}                
  \subfloat{\includegraphics[width=90mm,height=67mm]{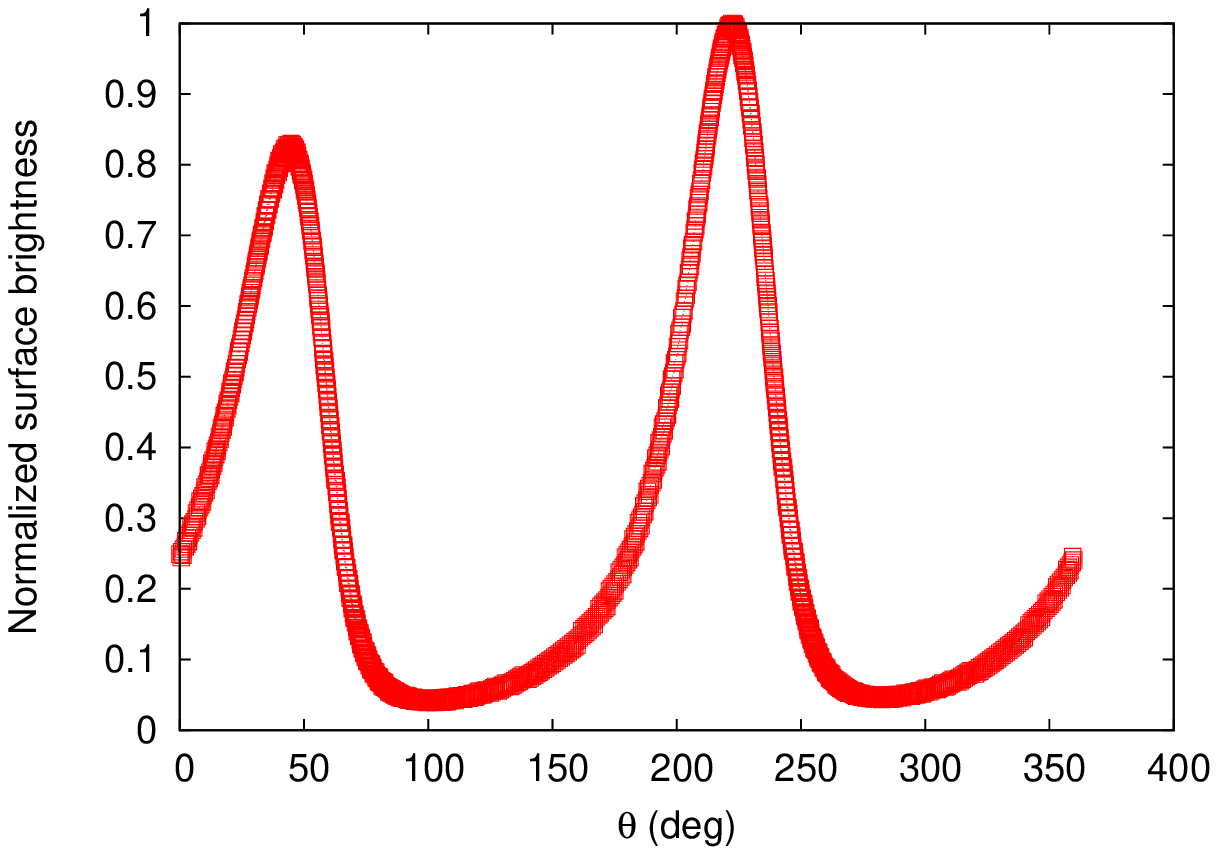}} \\           
  \caption{Result for the simulation of a $4.2$~M$_{\rm J}$ planet with $a_{p}=97$~AU, $e_{p}=0.06$ with a disc of multiple grain sizes, $s=8,100$ and $1000$~$\mu$m. (left) synthetic image at 70~$\mu$m roughly aligned with the \textit{Herschel} image, (right) normalized azimuthal surface brightness profile at 100 $\mu$m for $r=135$~AU: a 15$\%$ discrepancy in the surface brightness between the north-west and south-east lobes of the disc.} 
\label{fig:7}  
\end{center}
\end{figure*}
With initial planet parameter, $e_{p}=0.06$ and $a_{p}=97$~AU, we found a brightness asymmetry on the $70$~$\mu$m of about 18$\%$ and of about 15$\%$ at $100$~$\mu$m, while the initial configuration with $e_{p}=0.05$ and $a_{p}=102$~AU led to a asymmetry of 13$\%$ at 70~$\mu$m and 10$\%$ at 100~$\mu$m. Figure~\ref{fig:7} shows the \textit{MCFOST} synthetic image at 70~$\mu$m of the model with $e_{p}=0.06$ and $a_{p}=97$~AU along with the azimuthal surface brightness profile of the disc for a projected radius of $r=135$~AU at 100~$\mu$m.
We conclude that our best fits, assuming a planet between $97<a_{p}< 102$~AU and $0.05<e_{p}<0.06$, are consistent with both the brightness asymmetry and eccentricity found in the \textit{Herschel} image by ~\cite{2012A&A...537A.110L}.\\

\subsection{Observed point source near HD~207129}
~\cite{2010AJ....140.1051K} noticed a bright point source in the coronographic image of HD~207129 located towards the south edge of the image at a projected distance from the star that we estimate to be 110 AU. Using the observed object's brightness and spectral evolutionary models, ~\cite{2010AJ....140.1051K} estimated a mass of $\sim$ 20~M$_{\rm J}$. Since no follow-up observations have yet been made, it is currently unknown if this bright object is a background source or is comoving with the system. To investigate, we ran an additional simulation with the initial conditions for the disc taken from Table~\ref{Table1}, and instead of using the parameters predicted by ~\cite{2014ApJ...780...65R}, we place a  20 M$_{\rm J}$ planetary companion at $a_{p}$= 225~AU (the deprojected semi-major axis we estimate from the observation) with $e_{p}=e=0.08$. We ran the simulation for the same duration as the other models ($t_{sim}=20t_{sec}$), however the simulation ended at 0.9 $t_{sim}$ due to the total disruption of the belt. \\

\begin{figure}
\begin{center}
   \includegraphics[width=90mm,height=67mm]{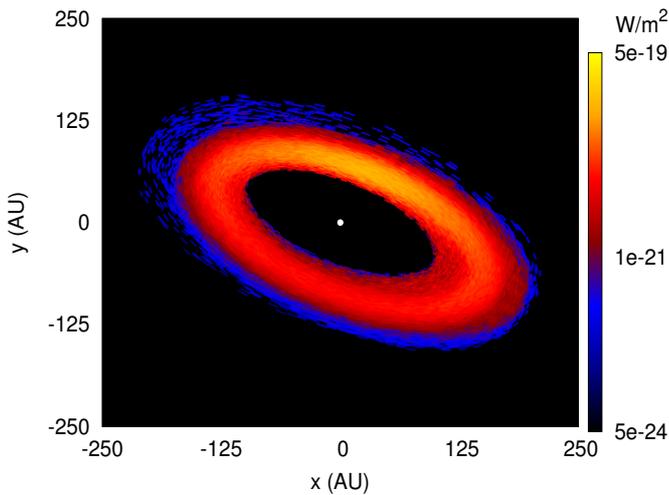}
  \caption{Synthetic scattered light image at 1.03 $\mu$m of our model for HD~207129 with an external planetary companion at $a_{p}$= 225~AU with $m_{p}$= 4.2~M$_{\rm J}$. The resulting disc structure is wider, with a larger offset and eccentricity than the best fit model presented in Figure \ref{fig:5} and the observations.} 
\label{fig:8}  
\end{center}
\end{figure}

Assuming the mass of the point source was highly overestimated (for example, due to accretion of dust around the planet, which dramatically increases the brightness), we repeated the simulation, but with a lower perturber mass $m_{p}$=4.2~M$_{\rm J}$ as estimated by ~\cite{2014ApJ...780...65R} with the same orbit as previously. At the end of the simulation we found that the resulting debris disc could be described as an ellipse located at 150 AU with $e$=0.18, $\delta=$15~AU and $\Delta_{r}=$40~AU -- see Figure~\ref{fig:8}. The corresponding $\Delta r/r_{0}$ ratio is 0.26. All parameters from the disc, producing $\chi^{2}=106.5$, are greater than those observed by ~\cite{2010AJ....140.1051K}. It therefore seems likely that this bright point source is a background source.

\subsection{Comparison with Rodigas et al.}
The predictions of Rodigas et al. aimed at constraining the maximum mass and minimum semi-major axis of a potential companion interior to a debris disc. To achieve this, they used N-body simulations with a broad range of planetary and disc parameters to derive an empirical equation linking the planetary parameters to the disc width. In this paper, we focus on modeling two specific debris discs to derive the mass, semi-major axis and eccentricity of the potential planet assuming the planet is sculpting the inner edge and eccentricity of the disc. 

Although we also use N-body simulations to model our planet-disc systems, a few differences with the method followed of Rodigas et al. should be pointed out. While Rodigas et al. includes radiation pressure acting on bound grains, we also include the stellar wind and PR drag acting on bound as well as unbound grains. As a result, small grains can be pushed outward by radiation pressure as well as dragged inwards by both PR and stellar wind drag. Therefore our simulations are expected to result in broader discs than those of Rodigas et al.

The Rodigas et al. simulations use a very specific initial setup up by constraining the initial disc eccentricity to be equal to that of the planet, forcing the disc and planet to be apse-aligned, and using an initially infinitesimally narrow parent body belt. We use a more relaxed configuration with a dynamically warm broad parent body belt with a range of initial eccentricities, and do not force the planet-disc alignment. In \cite{2015arXiv150908589T}, we addressed the issue of using different initial conditions in N-body debris discs simulations. We found that simulations in which the disc is initially secularly forced in eccentricity and apse aligned with the planet produce (i) narrower discs than when using an initial dynamically warm disc, and (ii) if, in addition to the forced initial conditions, an initially narrow is used, the final peak emission location is shifted outward. Therefore, since narrow forced initial conditions overestimate the distance between the planet and the disc peak emission, it is not surprising that our best fit models required a planet with a larger semi-major axis as well as a smaller mass than predicted by Rodigas et al.

Finally, Rodigas et al.  integrate the trajectory of their parent bodies for about 1000 orbits before running their simulations for the test particles for about 1 $t_{coll}$. Our simulations directly integrate the trajectories of the test particles for 20 $t_{sec}$, which corresponds to 19 $t_{coll}$ for HD~202628 and 23.5 $t_{coll}$ for HD~207129. By running the simulations for a longer duration, the particles have more time to be diffused and thus result a broader final configuration.

\subsection{Detectability of predicted planets}
The new generation of  direct exoplanet imagers, such as the Gemini Planet Imager (GPI) and the Spectro-Polarimetric High-Contrast Exoplanet Research (SPHERE), are expected to reach contrast levels (defined as the ratio of the planet to star brightness) of $10^{-7}$ to  $10^{-8}$ in scattered light and $10^{-6}$ in infrared within $\sim 6 \lambda/D$ of the star, where $D$ the diameter of the telescope aperture \citep{2010exop.book..111T}. Both predicted companions around HD~202628 and HD~207129 have a maximum angular separation of 5 and 6.3 arcsec respectively to their host stars. Such a planet-star separation is greater than the minimal distance ($\sim 6 \lambda/D$) imposed by diffraction, but also leads to a very low contrast level of $\sim 7 \times 10^{-12}$ in the visible spectrum, assuming a planet radius of 1.5 $R_{J}$ and the geometric albedo of Jupiter. In the near IR, giant planets are self-luminous and their important thermal emission can lead to a contrast several orders of magnitude brighter than in the optical, especially for young planetary systems. Extrapolating from the hot star evolution model of \cite{2008ApJ...683.1104F} and assuming a 4 Jupiter mass planet orbiting a Gyr old star, the expected absolute magnitude for such planet is $\sim 22$. Given the magnitude of both HD~202628 and HD~207129, this would correspond to a $\sim 3.5 \times 10^{-8}$ contrast in the H band, which still remains below the current detection threshold. Therefore, even with recent instrumental improvements, the direct detection of both potential companions in a near future unfortunately remains unlikely.

\section{Conclusions}
In order to study the perturbations induced by a planet on the structure of a debris disc, we have modified an N-body integrator to include stellar radiation pressure, PR drag and stellar wind. To compare the results of our numerical simulations with  observations, we used the 3D Monte Carlo radiative transfer code \textit{MCFOST} to produce synthetic images of the resulting debris discs. Our aim was to investigate the planetary configuration predicted by ~\cite{2014ApJ...780...65R} using the scattered light observations of the width of the debris discs detected around HD~207129 and HD~202628. We report the following results and conclusions:
\begin{itemize}
\item We found good agreement between our synthetic observations and scattered light observations for HD~202628 using a slightly higher planet eccentricity ($e_{p}=0.2$), a larger semi-major axis ($a_{p}=101$~AU) and a planetary mass five times smaller ($m_{p}=3$~M$_{\rm J}$) than predicted by~\cite{2014ApJ...780...65R}. Our best fit is in good agreement with the planet location estimated by ~\cite{2013AAS...22114414S}, \cite{2014MNRAS.443.2541P} and \cite{2015ApJ...798...83N}.
\item We found good agreement between our synthetic observations from numerical simulations and scattered light observations for HD~207129 using a planet located between $97<a_{p}<102$ AU, which is $5-10$~AU beyond the predictions of ~\cite{2014ApJ...780...65R}, with an slightly smaller eccentricity ($0.05<e_{p}<0.06$). We demonstrate that such a planet can explain the brightness asymmetries detected in  the far-infrared with \textit{Herschel/PACS} by ~\cite{2012A&A...537A.110L}.
\item We suggest that the observed bright point source in the southern corner of the coronographic image of HD~207129 is likely a background source. Numerical simulations showed that if this object was part of the system, the resulting debris disc structure would not match the current observed structure.
\end{itemize}
Overall, we conclude that the predictions of ~\cite{2014ApJ...780...65R} provide a good starting point for estimating the orbit of a potential planetary companion to debris discs, but should be complemented by numerical studies to test if the predicted planet can indeed induce the observed features in the disc.

\section*{Acknowledgements}
This work was performed on the swinSTAR supercomputer at Swinburne University of Technology. We thank Hal Levison for his assistance with \textit{SWIFT}, Chris Blake for his assistance with the fitting procedure, Christophe Pinte for his help with \textit{MCFOST} and Thayne Currie for his advices on planet detectability. We also thank Alexander Mustill for his suggestions to improve this paper. E.T. acknowledges the support of a Swinburne University Postgraduate Research Award (SUPRA).

\begin{appendix}
\section{Simulation duration}
The lifetime of grains in the disc is determined by three separate physical processes: Poynting-Robertson (PR) drag and radiation pressure which both tend to remove grains from the system, as well as destructive collisions between grains, which also reduce the population. \cite{2012AJ....144...45K} estimated that the collisional lifetime for grains in HD~202628 is about 100 shorter than the PR lifetime. Following \cite{1999ApJ...527..918W}, the collisional lifetime, $t_{coll}$, of a grain orbiting at a heliocentric distance $r$ is:
\begin{equation}
t_{coll} \sim \frac{t_{orb}}{4\pi\tau},
\end{equation}
where $t_{orb}$ is the orbital period of the grain at distance $r$, and $\tau$ is the effective optical depth of the disc. The peak of emission in the disc for HD 207129 and HD 202628 is located around 163 and 180 AU respectively. \cite{2012AJ....144...45K} estimated the optical depth, $\tau$, for both discs to be $\sim 2 \times10 ^{−4}$, therefore for grains located near to the emission peak the corresponding collisional lifetime is $\sim$ 940,000 years for HD 202628 and $\sim$ 790,000 years for HD~207129. 

Because destructive collisions are said to be the main destroyer of grains on a very short timescale, it has been argued by \cite{2009ApJ...707..543S} and \cite{2012A&A...547A..92T} that collisions can prevent planet-induced structures forming in the disc. More recently, \cite{2014arXiv1409.7609K} showed that asymmetries resulting from stochastic collisions are erased $t_{coll}$ after their formation due to radiation pressure redistributing the orbits of small grains. Therefore it is unclear that duration of a simulation should be limited by $t_{coll}$. This topic requires a complete study that is beyond the scope of this article. 

In this study secular perturbations are the crucial interaction between the planet and the disc. These perturbations impose the eccentricity of the disc to precess about a forced value on a timescale $t_{sec}$ given by:
\begin{equation}
t_{sec}=\frac{2\pi}{A \times t_{year}},
\end{equation} 
where $t_{year}$ is the number of seconds in a year and $A$ is the precession rate as a function of the planet $a_{p}$, $m_{p}$ and disc semi-major axis, $a$, and mean motion, $n$ (see \cite{1999ApJ...527..918W} for a complete expression of $A$). As reported by \cite{2009ApJ...693..734C}, the forced eccentricity tends toward the value of the planetary eccentricity, $e_{p}$, while $a_{p}$ tends toward $a$. Using the location of the peak emission as the disc semi-major axis, we estimate $t_{sec}$ for the planetary configuration predicted by \cite{2014ApJ...780...65R} to be 893,340 years for HD~202628 and 972,450 years for HD~207129.

These different physical processes need to be taken into account when setting the simulation duration. To test the impact of the final integration time on the resulting disc structure, we ran several simulations for the Rodigas et al. predicted model of HD~207129 with different simulation durations: $2,5,10,20,40,100$ and $150~t_{sec}$. We then created synthetic images and fit the disc parameters of each simulation -- see Table~\ref{Table5} for the results.

\begin{table}
\renewcommand{\arraystretch}{1.0}
\caption{Disc parameters for simulations of HD~207129 with different durations using the initial planetary parameters predicted by Rodigas et al.}
\label{Table4}
\centering
\begin{tabular}{|c|c|c|c|c|}
\hline
$t_{sim}$ ($t_{sec}$) &$t_{sim}$ ($t_{coll}$)& $\delta$ (AU) & $r_{peak}$ & $\Delta r/r_{0}$ \\
\hline
2 & 2.5 &  13.3 & 162.5 & 0.187 \\
5 & 6.1 &	13.4  &	161.0 &	0.195  \\
10 & 12  &	12.3  &	159.1 &	0.195 \\
20 & 25  &	11.1  &	156.0 &	0.183 \\
40 & 50 &	10.3  &	154.9 &	0.185  \\
100 & 120 &	9.6   &	154.7 &	0.184  \\
150 & 180 &	10.5  &	155.1 &	0.189  \\
\hline
\end{tabular}
\end{table}
We find that variations in the disc offset, width and peak location rapidly tend to constant values (less than $10\%$ variation) after $25~t_{coll}$, which is about 19,500,000 years. The corresponding secular timescale for this duration is $20~t_{sec}$. We therefore chose to run each simulation for a time corresponding to $20~t_{sec}$ with a timestep, $\Delta t$, taken to be a $1/30$ of the planet period, since no strong variation in the disc offset, width and peak location are noticeable after this period.

\section{Recording time of test particles during the simulation}
We ran two simulations to check that our results are not affected by our choice of initial recording time: one with the initial data recording time set to $t_{init}=0.0$ (meaning that test particle and planet positions are recorded from the beginning of the simulation), and one with $t_{init}=t_{sec}$, where $t_{sec}$ is the secular time of the planet. We created synthetic images and fit the disc parameters for both simulations and found the same final disc parameters to within $4\%$. We conclude that the initial conditions have little effect on the final disc parameters. We therefore set the initial recording time $t_{init}=0.0$ for the rest of the study.

Following a similar method, we also tested that the final fit of the disc parameters is independent of the output recording time, $t_{dump}$. We ran two simulations: first with $t_{dump}=4/3~P_{p}$ (which corresponds to $\sim 800$ years for both systems) and also with $t_{dump}= t_{coll}/100 \sim 8000$ years. We found that both simulations resulted in similar disc parameters to within $7\%$, and we again conclude that the determination of the disc parameters in independent of $t_{dump}$. We use $t_{dump}$ equal to $4/3~P_{p}$ for the rest of the study.

\begin{table*}
\caption{Simulation input planetary parameters ($m_{p}, a_{p}$ and $e_{p}$) and the resulting disc structure ($\delta, e, r_{0}$ and $\Delta r/r_{0}$) for HD~202628. A total of 150 simulations were run to perform this study, however for clarity purposes only the best fit (BF) values are given within a set of simulations for a planetary mass of 15.4, 9.9, 4.0 and 2.0 M$_{\rm J}$, along with all simulations with the planetary mass equal to the best fit value of 3~M$_{\rm J}$. The last column represents the total $\chi_{tot}^{2}$ for the model.}
\label{Table5}
\begin{tabular}{ccccccccc}
\hline
Model & $m_{p}$ (M$_{\rm J}$) & $a_{p}$ (AU) & $e_{p}$  &  $\delta$ (AU)& $e$ & $r_{0}$ (AU) & $\Delta r/r_{0}$  & $\chi_{tot}^{2}$\\
\hline
Rodigas & 15.4 & 71 & 0.18 & 11.8 & 0.08 &  157.4 & 0.41 & 2088.0 \\ 
BF & 15.4  & 111  & 0.26 & 23.3 &  0.11 &  219.2 & 0.45 & 309.2   \\
BF & 9.9 & 101 & 0.18 & 27.4 &  0.14 &  192.4 &  0.27 &  15.3 \\
BF & 4.0 & 111 & 0.18 & 27.4 &  0.15 &  181.6 & 0.36 & 14.9\\ 
BF & 2.0 & 101 & 0.26 & 28.2 &  0.16 &  175.5 & 0.37 & 3.7 \\
1 & 3.0 & 71 & 0.16 & 6.0 & 0.04 & 148.8 & 0.39 & 3423.7 \\
2 & 3.0 & 71 & 0.18 & 6.8 & 0.05 & 144.6 & 0.39 & 3221.7\\
3 & 3.0 & 71 & 0.20 & 8.0 & 0.05 & 146.4 & 0.38 & 2875.6 \\
4 & 3.0 & 71 & 0.22 & 8.4 & 0.06 & 143.9 & 0.38 & 2779.7\\
5 & 3.0 & 71 & 0.24 & 9.5 & 0.06 & 146.4 & 0.38 & 2494.5 \\
6 & 3.0 & 71 & 0.26 & 9.8 & 0.07 & 145.4 & 0.37 & 2424.9\\
7 & 3.0 & 81 & 0.16 & 14.5 & 0.09 & 159.1 & 0.38 & 1328.7 \\
8 & 3.0 & 81 & 0.18 & 15.8 & 0.10 & 158.1 & 0.37 & 1101.4\\
9 & 3.0 & 81 & 0.20 & 17.8 & 0.11 & 155.6 & 0.37 & 804.6\\
10 & 3.0 & 81 & 0.22 & 18.5 & 0.12 & 156.8 & 0.38 & 711.5\\
11 & 3.0 & 81 & 0.24 & 19.7 & 0.13 & 155.7 & 0.37 & 556.2\\
12 & 3.0 & 81 & 0.26 & 20.0 & 0.13 & 156.0 & 0.36 & 523.0\\
13 & 3.0 & 91 & 0.16 & 20.4 & 0.13 & 159.5 & 0.40 & 465.4\\
14 & 3.0 & 91 & 0.18 & 22.4 & 0.13 & 167.6 & 0.37 & 254.7\\
15 & 3.0 & 91 & 0.20 & 24.2 & 0.14 & 169.6 & 0.37 & 131.9\\
16 & 3.0 & 91 & 0.22 & 25.8 & 0.15 & 167.3 & 0.36 & 64.2\\
17 & 3.0 & 91 & 0.24 & 26.9 & 0.16 & 165.9 & 0.35 & 36.0\\
18 & 3.0 & 91 & 0.26 & 28.1 & 0.17 & 164.7 & 0.36 & 25.5 \\
19 & 3.0 & 101 & 0.16 & 23.1 & 0.13 &  173.7 & 0.35 & 193.8 \\
20 & 3.0 & 101 & 0.18 & 26.9 & 0.15 &  178.8 & 0.34 & 16.7 \\
21 & 3.0 & 101 & 0.20 & 28.5 & 0.16 &  179.3 & 0.34 & 0.78 \\
22 & 3.0 & 101 & 0.22 & 29.6 & 0.16 &  179.5 & 0.34 &  9.9 \\
23 & 3.0 & 101 & 0.24 & 33.2 & 0.18 &  181.6 & 0.33 & 154.3 \\
24 & 3.0 & 101 & 0.26 & 34.0 & 0.19 &  180.6 & 0.32 & 213.3\\
25 & 3.0 & 111 & 0.16 & 26.5 & 0.15 &  181.8 & 0.29 & 24.2\\
26 & 3.0 & 111 & 0.18 & 27.8 & 0.15 &  180.5 & 0.31 & 2.5 \\
27 & 3.0 & 111 & 0.20 & 30.8 & 0.17 &  183.1 & 0.32 & 39.7 \\
28 & 3.0 & 111 & 0.22 & 31.7 & 0.18 &  178.6 & 0.32 & 72.9\\ 
29 & 3.0 & 111 & 0.24 & 34.4 & 0.18 &  188.8 & 0.33 & 242.5\\
30 & 3.0 & 111 & 0.26 & 35.9 & 0.20 &  179.3 & 0.32 & 374.8\\
\hline
\end{tabular}
\end{table*}

\begin{table*}
\caption{Simulation input planetary parameters ($m_{p}, a_{p}$ and $e_{p}$) and the resulting disc structure ($\delta, e, r_{0}$ and $\Delta r/r_{0}$) for HD~207129. A total of 40 simulations were run for this study. The last column represents the total $\chi_{tot}^{2}$ for the model.}
\label{Table6}
\begin{tabular}{ccccccccc}
\hline
Model & $m_{p}$ (M$_{\rm J}$) & $a_{p}$ (AU) & $e_{p}$  &  $\delta$ (AU)& $e$ & $r_{0}$ (AU) & $\Delta r/r_{0}$  & $\chi_{tot}^{2}$\\
\hline
Rodigas &  4.2 & 92 & 0.08 & 11.1 &  0.07 & 156.1 &  0.18 & 10.0\\
2 &  4.2 & 92 & 0.07 & 9.7 &  0.06 &  157.1 &  0.17 & 7.3 \\
3 &  4.2 & 92 & 0.06 & 8.6 &  0.05 &  157.6  & 0.17 & 6.2 \\
4 &  4.2 & 92 & 0.05 & 6.4 &  0.04 &  158.1 &  0.17 & 5.0 \\
5 &  4.2 & 92 & 0.04 & 6.1 &  0.04 &  153.5 &  0.2 &  18.9 \\
6 &  4.2 & 92 & 0.03 & 6.3 &  0.04 &  151.4 &  0.19 &  28.0\\
7 &  4.2 & 92 & 0.02 & 3.9 &  0.02 &  150.1 &  0.18 &  34.7 \\
8 &  4.2 & 92 & 0.01 & 1.9 &  0.01 &  149.5 &  0.16 & 38.4 \\
9 & 4.2 & 97 & 0.08 & 10.6 &  0.06 &  162.0 &  0.21 & 0.30 \\
10 & 4.2 & 97 & 0.07 & 9.3 &  0.06 &  156.5 &  0.20 &  9.0 \\
11 & 4.2 & 97 & 0.06 & 9.7 &  0.06 &  163.0 &  0.18 & 0.07 \\
12 & 4.2 & 97 & 0.05 & 10.9 & 0.07 &  162.1 &  0.16 & 0.21 \\
13 & 4.2 & 97 & 0.04 & 7.4 & 0.04 &   159.0 &  0.15 & 3.4 \\
14 & 4.2 & 97 & 0.03 & 5.9 &  0.04 &  160.0 &  0.13 & 2.2 \\
15 & 4.2 & 97 & 0.02 & 3.7 & 0.02 &   158.7 &  0.12 & 4.2 \\
16 & 4.2 & 97 & 0.01 & 1.9 &  0.01 &  157.7 &  0.11 & 6.4\\
17 & 4.2 & 102 & 0.08 & 11.4 & 0.07 & 160.8 &  0.24 & 1.2\\
18 & 4.2 & 102 & 0.07 & 9.7 & 0.06 &  161.4 &  0.21 & 0.61 \\
19 & 4.2 & 102 & 0.06 & 8.6 & 0.05 &  161.6 &  0.19 & 0.44 \\
20 & 4.2 & 102 & 0.05 & 7.3 & 0.04 &  162.2 &  0.17 & 0.16 \\
21 & 4.2 & 102 & 0.04 & 6.4 & 0.03 &  163.3 &  0.15 & 0.14 \\
22 & 4.2 & 102 & 0.03 & 5.4 & 0.03 &  164.9 &  0.13 & 1.0\\
23 & 4.2 & 102 & 0.02 & 4.6 & 0.03 &  165.4 &  0.11 & 1.8 \\
24 & 4.2 & 102 & 0.01 & 2.8 & 0.02 &  165.1 &  0.10 & 1.6 \\
25 & 4.2 & 107 & 0.08 & 16.3 & 0.10 & 157.2 &  0.26 & 121.8 \\
26 & 4.2 & 107 & 0.07 & 16.5 & 0.10 & 158.2 &  0.25 & 137.5\\
27 & 4.2 & 107 & 0.06 & 14.5 & 0.09 & 161.4 &  0.25 & 24.2 \\
28 & 4.2 & 107 & 0.05 & 15.5 & 0.1 &  160.0 &  0.24 & 69.8 \\
29 & 4.2 & 107 & 0.04 & 14.5 & 0.09 & 161.6 &  0.24 & 23.6 \\
30 & 4.2 & 107 & 0.03 & 7.7 & 0.04 &  166.9 &  0.22 & 3.4\\
31 & 4.2 & 107 & 0.02 & 6.0 & 0.04 &  168.9 &  0.22 & 7.5\\
32 & 4.2 & 107 & 0.01 & 3.5 & 0.02 &  169.7 &  0.20 &  9.5\\
33 & 4.2 & 112 & 0.08 & 10.7 & 0.07 & 152.6 &  0.15 & 23.0 \\
34 & 4.2 & 112 & 0.07 & 10.2 & 0.07 & 153.6 &  0.16 & 18.8 \\
35 & 4.2 & 112 & 0.06 & 8.2 & 0.05 &  153.8 &  0.14 & 18.1 \\
36 & 4.2 & 112 & 0.05 & 7.3 & 0.05 &  154.5 &  0.14 & 15.5 \\
37 & 4.2 & 112 & 0.04 & 5.7 & 0.04 &  155.9 &  0.13 & 10.9\\
38 & 4.2 & 112 & 0.03 & 4.0 & 0.03 &  155.8 &  0.13 & 11.1\\
39 & 4.2 & 112 & 0.02 & 3.1 & 0.02 &  155.4 &  0.12 & 12.5\\
40 & 4.2 & 112 & 0.01 & 1.7 & 0.01 &  154.9 &  0.14 & 13.8\\
\hline
\end{tabular}
\end{table*}

\end{appendix}

\end{document}